\documentclass[10pt,letterpaper]{article}
\usepackage[top=0.85in,left=2.75in,footskip=0.75in]{geometry}

\usepackage{amsmath,amssymb}

\usepackage{changepage}

\usepackage{textcomp,marvosym}

\usepackage{cite}

\usepackage{nameref,hyperref}

\usepackage[right]{lineno}

\usepackage[nopatch=eqnum]{microtype}

\usepackage[table]{xcolor}

\usepackage{array}

\newcolumntype{+}{!{\vrule width 2pt}}

\newlength\savedwidth

\raggedright
\usepackage[aboveskip=1pt,labelfont=bf,labelsep=period,justification=raggedright,singlelinecheck=off]{caption}

\makeatletter
\renewcommand{\@biblabel}[1]{\quad#1.}
\makeatother

\usepackage{lastpage,fancyhdr,graphicx}
\usepackage{epstopdf}
\fancyheadoffset[L]{2.25in}
\fancyfootoffset[L]{2.25in}
\begin{document}
\vspace*{0.2in}

\begin{flushleft}
{\Large
\textbf{Simple models outperform single-cell foundation models on downstream benchmarks}
}
\newline
\\
Huan Souza\textsuperscript{1*},
Pankaj Mehta\textsuperscript{1, 2}
\\
\bigskip
\textbf{1} Department of Physics, Boston University, Boston, MA, 02215, USA
\\
\textbf{2} Faculty of Computing and Data Science, Boston University, \\Boston, MA, 02215, USA
\\
\bigskip

* Corresponding author\\
E-mail: hsouza@bu.edu (HS)
\end{flushleft}


%
\section*{Abstract}
Single-cell RNA sequencing (scRNA-seq) data exhibit strong and reproducible statistical structure. This has motivated the development of large-scale foundation models,  such as TranscriptFormer, that use transformer-based architectures to learn a generative model for gene expression by embedding genes into a latent vector space. These embeddings have been used to obtain state-of-the-art (SOTA) performance on downstream tasks such as cell-type classification, disease-state prediction, and cross-species learning.  Here, we ask whether similar performance can be achieved without utilizing computationally intensive deep learning-based representations. Using simple, interpretable pipelines that rely on careful normalization and linear methods, we obtain SOTA or near SOTA performance across multiple benchmarks commonly used to evaluate single-cell foundation models, including outperforming foundation models on out-of-distribution tasks involving novel cell types and organisms absent from the training data. Our findings highlight the need for rigorous benchmarking and suggest that the biology of cell identity can be captured by simple linear representations of single cell gene expression data.

\clearpage
\newgeometry{top=0.85in,left=1in,right=1in,footskip=0.75in}

\section*{Introduction}
Advances in single-cell transcriptomics have transformed our ability to discover and experimentally characterize cell types across tissues and organisms ~\cite{musser2021profiling, massri2021developmental, cole2024updated, briggs2018dynamics, wagner2018single, tabula2020single, shahan2022single, czi2025cz, rood2025human, stuart2019comprehensive, butler2018integrating, cao2019single, hao2021integrated, tarashansky2021mapping}. A major achievement of this research program has been the creation of  ``cell atlases'' that catalogue cellular gene expression with single cell resolution \cite{quake2022decade,rood2025human}. Cell atlases often contain data from hundreds of million cells, with the amount of data expected to grow as sequencing costs continue to drop and new experimental techniques and modalities emerge \cite{hemberg2025insights}. For this reason, there is immense interest in leveraging single cell RNA-sequencing (scRNA-seq) data to extract biological insights about the molecular basis of cellular identity.

Although every cell in an organism is genomically identical, different cell types express different subsets of genes, giving rise to observed phenotypic and functional differences. One seductive idea is that the essential aspects of cellular identity are encoded in the statistical properties of cellular gene expression profiles. If true, this suggests that learning good representations of single-cell data offers a powerful route for understanding cellular identity across cell types, tissues, and species. This perspective has motivated a broad range of statistical approaches for analyzing scRNA-seq datasets.

\begin{figure}[!htbp]
     \includegraphics[width=0.9\linewidth]{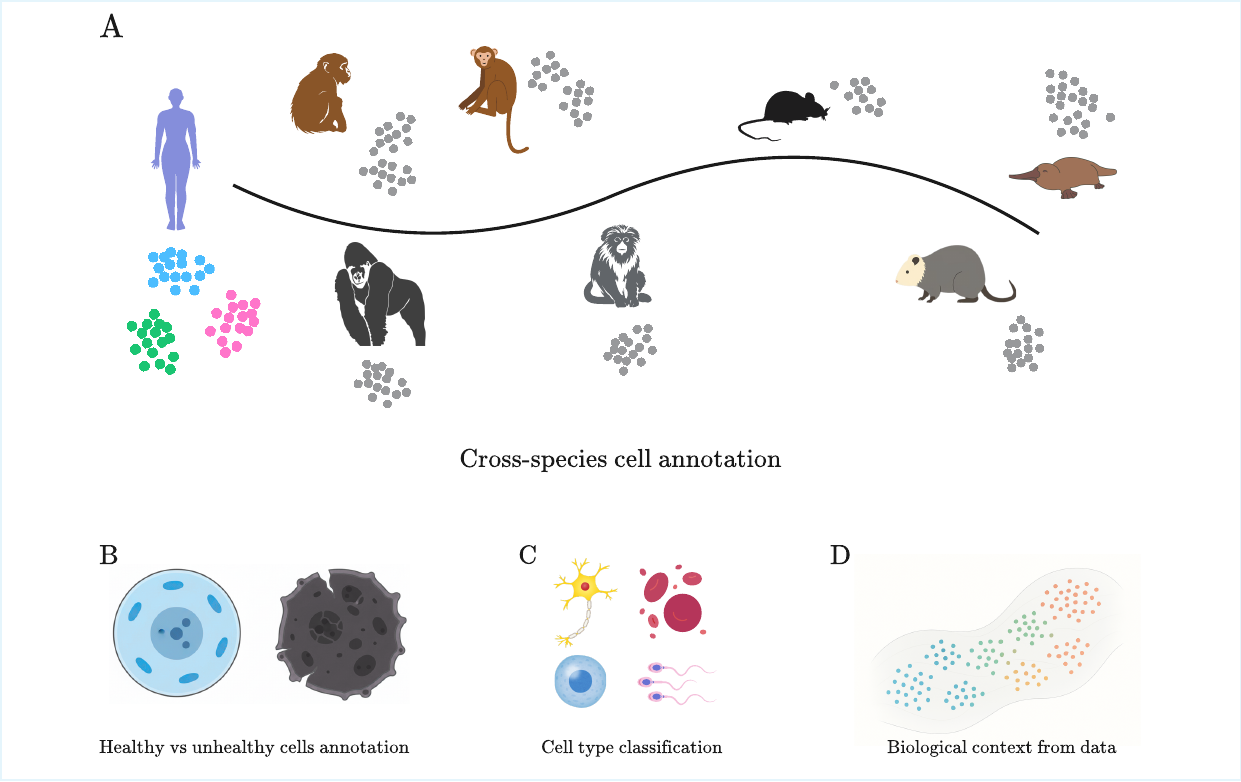}
     \caption{{\bf Downstream Tasks/Benchmarks analyzed in this paper}. \textbf{A}. Cross species cell annotation. The goal of this task is to use labeled cells from one species (e.g. humans) to annotate cell types in another species. \textbf{B}. Discrimination between healthy and infected cells. \textbf{C}. Cell type classification. \textbf{D}. Extracting biological context (i.e. gene-TF interactions) from data.}
     \label{fig1}
 \end{figure}

The hope that complex statistical models can be used to learn new biology is buttressed by the tremendous success of protein language models at tasks such as sequence-structure prediction (see \cite{bjerregaard2025foundation,weissenow2025protein} for recent reviews). Motivated in part by this work, several works have trained large self-supervised “foundation models” on cell atlas data
\cite{bian2024scmulan, hao2024large, cui2024scgpt, yang2024genecompass, schaar2024nicheformer, kalfon2025scprint, rosen2023universal, pearce2025cross}. Models such as TranscriptFormer use transformer-based architectures to learn a generative model for gene expression (i.e. mRNA counts) by embedding genes into a latent vector space. Gene embeddings from foundation models  have been used to obtain state-of-the-art (SOTA) performance on downstream tasks such as cell-type classification, disease-state prediction, and cross-species learning. This is often cited as evidence that models like TranscriptFormer learn general-purpose biological representations of gene expression that can serve as inputs for more ambitious “virtual cell” models \cite{bunne2024build}.

	Still, it remains unclear if the representations learned by foundation models capture biological structure beyond that which is already present in appropriately processed scRNA-seq data \cite{CZbenchmarks, fahsbender2025benchmarking}. A growing body of work has shown that simple, interpretable methods can perform remarkably well across diverse single-cell analysis tasks \cite{boiarsky2024deeper, bendidi2024benchmarking, kedzierska2025zero, csendes2025benchmarking, ahlmann2025deep, wenteler2024perteval, wu2025biology, atti2025fundamental, zhou2026benchmarking, xu2026cellbench, han2026benchmarking, hou2026unified, kendiukhov2026systematic, smith2020standard}. For example, linear and physics-inspired approaches such as Single-cell Type Order Parameters (scTOP) enable accurate cell-type classification, interpretable visualization of developmental dynamics, and principled analyses of cell fate transitions without requiring large-scale model training \cite{yampolskaya2023sctop, burgess2024generation, herriges2023durable, golden2025order, yampolskaya2025finding, alber2026bidirectional}. These observations raise two basic but under-explored questions: how complex is the structure of scRNA-seq data itself? What level of representational sophistication is required to extract the biologically relevant variation captured by current benchmarks?

	There are several good reasons to believe that embeddings from single cell foundation models may be less powerful than those from protein language models \cite{mehta2024twenty,chari2023specious}. In contrast to protein sequences, which are discrete, high-quality, and tightly constrained by biophysics, scRNA-seq data are sparse, noisy, and characterized by substantial technical variability, including dropout effects and batch-specific artifacts \cite{hicks2018missing}. Moreover, cellular identity reflects a complex interplay between gene expression, signaling, environmental context, and post-transcriptional regulation, rather than being fully specified by transcript counts alone. This is in stark contrast with proteins where the information needed to determine its three-dimensional structure is contained almost entirely in its sequence (Anfinsen's principle) \cite{anfinsen1973principles,lockless1999evolutionarily,socolich2005evolutionary,morcos2011direct}.

	Inspired by these considerations, we systematically compared the performance of simple, interpretable pipelines to that of large scale single-cell foundation models, evaluated with identical data splits on common downstream tasks (see Fig~\ref{fig1}) \cite{CZbenchmarks}. We find that by carefully choosing pre-processing steps and normalization procedures, it is possible to achieve SOTA or near-SOTA performance using simple representations where cells are viewed as vectors in gene expression space. The performance of these pipelines often exceeds that of foundation models, despite the fact that they require orders-of-magnitude less computational resources and have only a few hyperparameters (see \nameref{S1_Text} for an extended comparison of computational cost).

	These results motivate a unifying interpretation: much of the biologically relevant structure present in current scRNA-seq benchmarks is already accessible through low-complexity linear representations. Consequently, widely used evaluation tasks primarily reflect intrinsic properties of the data manifold rather than the discovery of new biologically meaningful structure (in this work, as is common in the machine learning literature, \ a \emph{manifold} refers to the effective low-dimensional structure formed by scRNA-seq data within the high-dimensional gene-expression space. This usage is informal and geometric, emphasizing distances, curvature, and separability). Our work highlights the need for more stringent evaluation protocols and provides strong evidence that simple representations in terms of normalized gene expression counts capture many of the key aspects of cell identity contained in scRNA-seq cell atlases.

\section*{Materials and methods}
\subsection*{scTOP package}
	For cross-species annotation, we used our previously published method, single-cell Type Order Parameter (scTOP)  \cite{yampolskaya2023sctop}. Moreover, the pre-processing function of scTOP was applied to every task in our work. scTOP quantifies cellular identity by projecting a cell's gene expression profile onto a defined cell type space. Each cell is processed independently to minimize batch effects.

	This preprocessing calculation is performed as follows. First, each cell's raw counts, $x_j$, are normalized by the total counts in that cell (for $j=1, \dots, G$):
	\begin{equation*}
		x_j^{\text{norm}} = \frac{x_j}{\sum_{k=1}^{G} x_k}
	\end{equation*}
	This normalized value is then transformed using a $\log_2(x+1)$ function:
	\begin{equation*}
		y_j = \log_2(x_j^{\text{norm}} + 1).
	\end{equation*}
	The $y_j$ value is ranked relative to all other transformed gene values in that cell, yielding $R_j$. The rank is converted to a percentile rank, $p_j$, using the convention:
	\begin{equation*}
		p_j = \frac{R_j}{G+1}
	\end{equation*}
	This percentile is then converted to a z-score, $S_j$, using the inverse of the standard normal cumulative distribution function (CDF), $\Phi^{-1}$:
	\begin{equation*}
		S_j = \Phi^{-1}(p_j)
	\end{equation*}
	The resulting $G$-dimensional vector, composed of the $S_j$ values, is used for the projection. Therefore, $S_j$ is the preprocessed single cell gene expression vector.

	Next, a reference basis is constructed from existing single-cell atlases to define the $C$ cell types of interest. For each of the $C$ cell types (e.g., Fibroblast, Macrophage, Basal), the gene expression profiles from a representative population (in this work, all training cells of that cell type) are first normalized individually (by total counts per cell, as above), and then averaged to create a single archetypal vector. This vector is then processed using the exact same $\log_2$ transformation, ranking, and z-scoring procedure (applied to the single averaged vector) to create the final reference vector, $\xi^{\mu}_i$, with $\mu = (1,..., C)$. These $C$ vectors form a non-orthogonal basis for the $C$-dimensional cell type space.

	The final scTOP scores, $a^{\mu}$, are calculated by projecting the query cell's vector (composed of $S_j$ values) onto this non-orthogonal basis. This is not a simple dot product but a de-correlated projection that accounts for the similarity between the reference types. First, an overlap matrix, $A_{\mu\nu}$, is computed, where each element is the dot product of two reference cell type vectors:
	\begin{equation*}
		A_{\mu\nu} = \sum_{j=1}^{G} \xi_{j}^{\mu} \xi_{j}^{v}.
	\end{equation*}
	The final score, $a^{\mu}$, for the query cell's similarity to cell type $\mu$ is a sum of the cell's similarity to all reference types, weighted by this inverse matrix:
	\begin{equation*}
		a^{\mu}=\sum_{v=1}^{C}\sum_{j=1}^{G}[A_{\mu\nu}]^{-1}\xi_{j}^{v}S_{j}.
	\end{equation*}
	The result is a set of $C$ scores for the query cell, representing its coordinates in the de-correlated cell type space. We then annotate the cell as the cell type with highest projection. In the cross-species benchmarks reported here, the reference vectors $\xi^{\mu}$ are built only from the training cells of each split, and each query cell is assigned to the cell type with the largest score $a^{\mu}$.

	\subsection*{ANOVA gene selection}

	For human cell annotation and disease state identification, after preprocessing samples using scTOP, we computed ANOVA F-statistic on the dataset, in order to select informative genes for each tissue. Let $x_{i,g}$ denote the standardized expression of gene $g$ in cell $i$, and let the dataset contain $C$ cell types with class means $\bar{x}_{g}^{(c)}$ and overall mean $\bar{x}_g$. The between-class variance is
	\[
	S_{B}(g) = \sum_{\mu=1}^{C} n_\mu \left( \bar{x}_{g}^{(\mu)} - \bar{x}_g \right)^{2},
	\]
	and the within-class variance is
	\[
	S_{W}(g) = \sum_{\mu=1}^{C} \sum_{i \in C_\mu} \left( x_{i,g} - \bar{x}_{g}^{(\mu)} \right)^{2}.
	\]
	The ANOVA score is
	\[
	F(g) = \frac{S_{B}(g)/(C-1)}{S_{W}(g)/(N-C)}.
	\]
	The $N_\text{ANOVA}$ genes with the highest values of $F(g)$ were retained ($N_\text{ANOVA}=22000$ for Tabula Sapiens and $N_\text{ANOVA}=20000$ for the COVID-19 dataset). This is computed in the training set. The selected genes during training are applied to the test set.

	\subsection*{Standardization}
	After ANOVA selection, each retained gene $g$ is standardized to zero mean and unit variance across cells,
	\begin{equation*}
		S'_{i,g} = \frac{S_{i,g} - \mu_g}{\sigma_g},
	\end{equation*}
	where $S_{i,g}$ is the expression of gene $g$ in cell $i$ after ANOVA selection, and $\mu_g$ and $\sigma_g$ are the mean and standard deviation of gene $g$ computed on the training cells only. The same $\mu_g$ and $\sigma_g$ are applied to the test cells.
	\subsection*{Principal Component Analysis}
	After ANOVA selection and standardization, Principal Component Analysis (PCA) is applied. PCA reduces the dimensionality of the data by identifying the primary axes of variance, which serves to capture the most significant biological signals while filtering out stochastic noise.

	Let $\mathcal{S}'$ denote the $N \times G'$ matrix of standardized, gene-selected values, where $N$ is the number of cells and $G'$ is the number of retained genes. The sample covariance matrix is
	\[
	\mathbf{\Sigma} = \frac{1}{N-1} \mathcal{S}'^\top \mathcal{S}'.
	\]
	PCA solves the eigenvalue problem to find the principal directions (eigenvectors) $\mathbf{u}_j$:
	\[
	\mathbf{\Sigma} \mathbf{u}_j = \lambda_j \mathbf{u}_j,
	\]
	where $\lambda_j$ is the eigenvalue representing the variance captured by $\mathbf{u}_j$. The principal directions are ordered such that $\lambda_1 \ge \lambda_2 \ge \dots \ge \lambda_{G'}$. By projecting a cell's standardized vector $\mathbf{s}'_i$ (a column vector representing the $i$-th cell's data from $\mathcal{S}'$) onto the first $d$ principal directions, we create a new, low-dimensional representation $\mathbf{z}_i$:
	\[
	\mathbf{z}_i = \mathbf{U}_d^\top \mathbf{s}'_i, \quad \text{where} \quad \mathbf{U}_d = (\mathbf{u}_1, \ldots, \mathbf{u}_d).
	\]
	This projection retains the majority of the structured variance (captured by the components with large eigenvalues) and discards the higher-order components (with small eigenvalues), which are more likely to represent noise. The resulting vector $\mathbf{z}_i$ is the final, noise-reduced representation of the cell. PCA is fit on the training cells only, and each component is whitened (scaled to unit variance) before classification.

	\subsection*{Logistic regression}

	For binary classification tasks (e.g., COVID-19 identification), we model the probability that cell $i$ (represented by its PCA vector $\mathbf{z}_i$) belongs to class $y_i=1$ as
	\[
	P(y_i = 1 \mid \mathbf{z}_i) = \sigma(\mathbf{w}^\top \mathbf{z}_i + b),
	\]
	where $\sigma(t) = 1/(1+e^{-t})$ is the logistic sigmoid. The parameters $(\mathbf{w},b)$ minimize the cross-entropy loss
	\[
	\mathcal{L} = -\sum_{i=1}^{N} \left[ y_i \log \sigma(\mathbf{w}^\top \mathbf{z}_i + b) + (1 - y_i)\log\bigl(1 - \sigma(\mathbf{w}^\top \mathbf{z}_i + b)\bigr) \right].
	\]

	For multi-class tasks with $K$ types, we use multinomial logistic regression. The probability that a cell belongs to class $k$ is
	\[
	P(y_i = k \mid \mathbf{z}_i) = \frac{\exp(\mathbf{w}_k^\top \mathbf{z}_i + b_k)}{\sum_{\ell=1}^{K} \exp(\mathbf{w}_\ell^\top \mathbf{z}_i + b_\ell)},
	\]
	and the loss is
	\[
	\mathcal{L} = -\sum_{i=1}^{N} \sum_{k=1}^{K} \mathbf{1}(y_i = k) \log P(y_i = k \mid \mathbf{z}_i).
	\]
	Here $\mathbf{1}(y_i = k)$ is an indicator function. In all tasks, the classifier uses class-balanced weights (each class is weighted inversely to its frequency in the training set), L2 regularization, and the L-BFGS solver, as implemented in scikit-learn.

	\subsection*{Leiden clustering}

	For the COVID-19 analysis, clustering was performed on the PCA-transformed data ($\mathbf{z}_i$ vectors). The Leiden algorithm maximizes the modularity
	\[
	Q = \frac{1}{2m} \sum_{i,j} \left( A_{ij} - \frac{k_i k_j}{2m} \right)\delta(c_i, c_j),
	\]
	where $A_{ij}$ is the adjacency matrix of the K-nearest-neighbor graph in PCA space, $k_i$ is the degree of node $i$, $m=\tfrac{1}{2}\sum_{i} k_i$, and $c_i$ denotes the cluster assignment of cell $i$. The algorithm iteratively maximizes $Q$ until convergence.

	Each cluster $C$ receives its own logistic regression model trained only on cells within $C$. A test cell is projected into PCA space, assigned to its nearest Leiden cluster, and evaluated by the corresponding classifier. We used 15 nearest neighbors and a resolution of 0.4, and test cells are assigned to clusters with a 15-nearest-neighbor classifier; clusters in which all training cells share the same label predict that label.

	\subsection*{Cosine similarity analysis}

	Cosine similarity between average profiles of two groups $A$ and $B$ is computed on the vectors after the preprocessing step from scTOP (before ANOVA or PCA) as
	\[
	\mathrm{cos}(A,B) = \frac{\mathbf{\mu}_A \cdot \mathbf{\mu}_B}{\|\mathbf{\mu}_A\| \,\|\mathbf{\mu}_B\|},
	\]
	where $\mathbf{\mu}_A$ and $\mathbf{\mu}_B$ are the mean vectors of the groups, using the full-dimensional scTOP-preprocessed vectors $\mathbf{S}_i$:
	\[
	\mathbf{\mu}_A = \frac{1}{|A|}\sum_{i\in A} \mathbf{S}_i,
	\qquad
	\mathbf{\mu}_B = \frac{1}{|B|}\sum_{i\in B} \mathbf{S}_i.
	\]
	(Here $\mathbf{S}_i$ is the $G$-dimensional vector $(S_{i,1}, \dots, S_{i,G})^\top$ for cell $i$.)

	\subsection*{Macro F1-score}

	For each class $k$, let precision and recall be
	\[
	\mathrm{Prec}_k = \frac{TP_k}{TP_k + FP_k},
	\qquad
	\mathrm{Rec}_k = \frac{TP_k}{TP_k + FN_k},
	\]
	where $TP_k$, $FP_k$, and $FN_k$ denote true positives, false positives, and false negatives, respectively. The F1-score for class $k$ is
	\[
	F1_k = \frac{2\, \mathrm{Prec}_k \, \mathrm{Rec}_k}{\mathrm{Prec}_k + \mathrm{Rec}_k}.
	\]
	The macro F1-score averages across classes:
	\[
	\mathrm{MacroF1} = \frac{1}{K}\sum_{k=1}^{K} F1_k.
	\]

	\subsection*{Foundation model baselines}
	To compare our pipelines with single-cell foundation models under identical conditions, we re-ran all foundation-model baselines ourselves instead of using previously reported numbers. Because extracting embeddings requires substantial GPU resources, we limited this comparison to TranscriptFormer (TF-Sapiens checkpoint) \cite{pearce2025cross}, ESM2-CE (as provided with TranscriptFormer) \cite{pearce2025cross}, UCE \cite{rosen2023universal}, a pretrained scVI reference model, and, for the cross-species task, Geneformer (gf-18L-316M-i4096 checkpoint). Embeddings were extracted zero-shot from the released pretrained models, without fine-tuning, starting from raw counts. For cross-species annotation, each species was embedded separately with all of its genes, following \cite{pearce2025cross}: TranscriptFormer and ESM2-CE use the ESM2 gene embeddings released with TranscriptFormer for each species, UCE uses its own protein embeddings for human, mouse, and rhesus macaque and the human orthologs of the remaining species, and the human-only models Geneformer and scVI use all genes with a human ortholog. For scVI, cells were mapped onto the pretrained reference model without retraining.

	For every task, the foundation-model embeddings (standardized for the cross-species and COVID-19 tasks, used as provided for Tabula Sapiens 2.0) were used as input to the same class-balanced logistic regression classifier described above, using exactly the same train-test splits as our pipelines. For cross-species annotation, cell types with at least 50 cells in the source species were split into stratified 80-20 training and test sets with five random seeds; within-species (diagonal) scores are computed on the held-out 20\%, and cross-species scores on all cells of the target species. For Tabula Sapiens 2.0, cell types with at least 100 cells in each tissue were split into stratified 80-20 training and test sets with five random seeds. For the COVID-19 task, we used ten-fold stratified cross-validation within each donor. In all cases, gene selection, standardization, PCA, and the classifier were fit on training cells only. Macro F1-scores include every class present in the training or test set, so that classes absent from training count as zero recall.

	\subsection*{Hyperparameters choice}\label{hyperparameters}

\begin{figure}[!htbp]
		\centering
		\includegraphics[width=\linewidth]{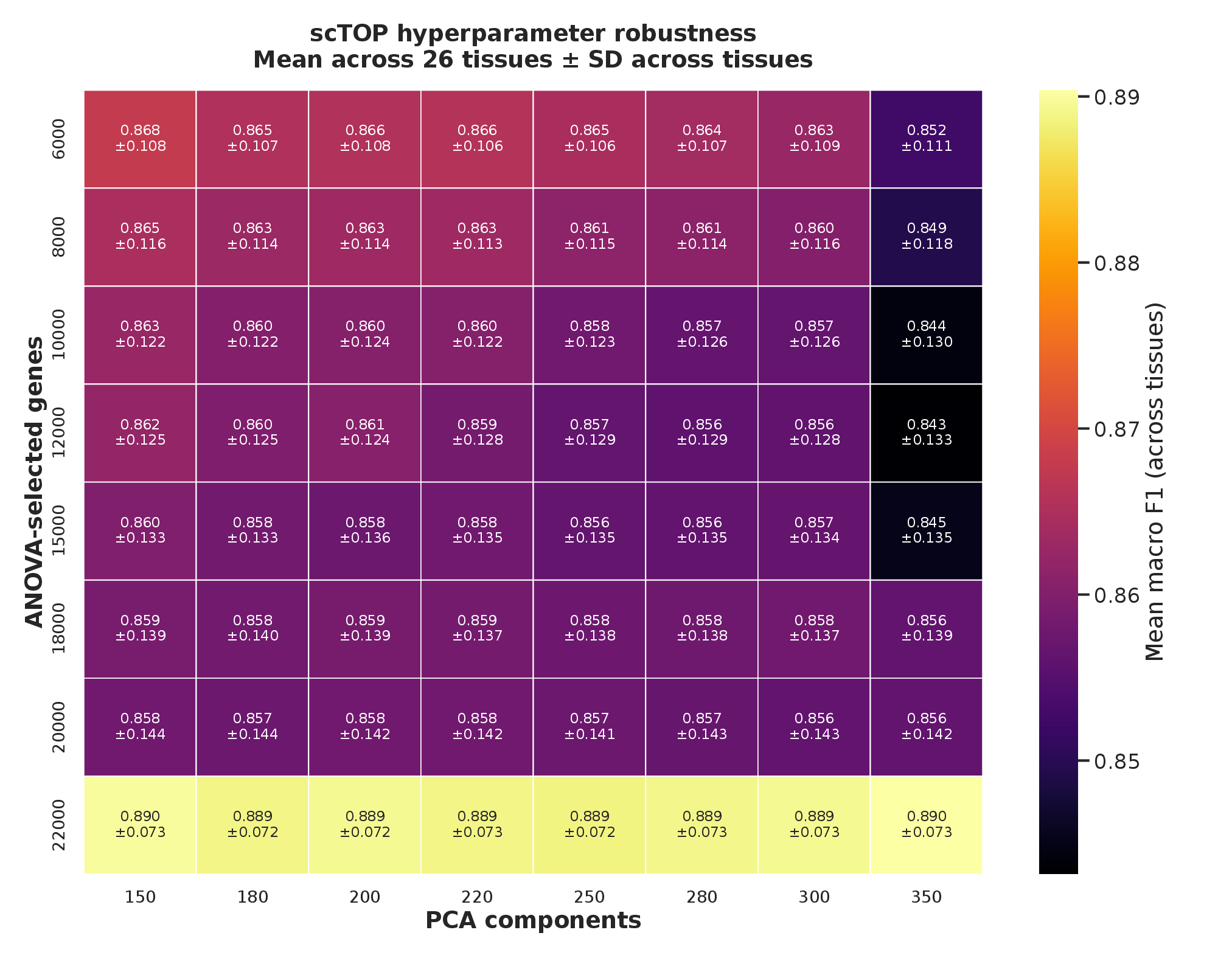}
		\caption{\textbf{Hyperparameters search:} Mean macro F1-scores for each choice of number of ANOVA selected genes and number of principal components. We notice that most choices lead to similar results.}\label{fig2}
	\end{figure}

	To choose the hyperparameter necessary for the classification task on the Tabula Sapiens dataset, we ran a grid search across different values of selected genes for ANOVA selection and number of PCA dimensions. For each choice, we evaluated three stratified 80-20 train-test splits. The results are shown in Fig~\ref{fig2}. As we can see, our results are robust across all choices. Then, we chose 22000 genes and 220 principal components as a way of keeping the maximum information without sacrificing performance.

	\subsection*{Gene--gene interaction analysis}

	All gene--gene interaction analyses were performed on scTOP-processed single-cell expression matrices, as described above, using the bone marrow cells of Tabula Sapiens 2.0 and the TFs E2F8, FOXM1, SPIB, MYBL2, and PTTG1. Let
	$\mathcal{S} \in \mathbb{R}^{N \times G}$ denote the processed expression matrix, where $G$ is the number of genes, $N$ the number of cells, and $s_{i g}$ the processed expression value of gene $g$ in cell $i$.

	For a given transcription factor (TF) $t$, TF activity was defined directly from its processed expression values. Cells were ranked according to $s_{i t}$, and those in the top quantile $q$ were designated as \emph{TF-active}. Formally, letting $Q_q(s_t)$ denote the empirical $q$-th quantile of the TF expression distribution, the set of TF-active cells was defined as
	\begin{equation}
		\mathcal{A}_t = \left\{ i \in \{1,\dots,N\} \mid s_{t i} > Q_q(s_t) \right\}.
	\end{equation}
	This procedure yields a binary conditioning variable indicating whether a cell exhibits high TF activity.

	To focus on co-occurrence structure rather than expression magnitude, each gene was modeled as a binary random variable indicating whether it is expressed in a given cell. Specifically, for each gene $g$ and cell $i$, we defined
	\begin{equation}
		Y_{i g} =
		\begin{cases}
			1, & \text{if } s_{i g} > 0, \\
			0, & \text{otherwise}.
		\end{cases}
	\end{equation}
	This binarization step reduces sensitivity to technical variability and mirrors gene-selection strategies employed in recent large-scale transcriptomic models.

	Genes expressed in fewer than a minimum fraction of cells were excluded from downstream analysis to avoid unstable estimates arising from extremely rare events.

	For each remaining gene $g$, we estimated its marginal probability of expression across all cells,
	\begin{equation}
		\hat{P}(g) = \frac{1}{N} \sum_{i=1}^{N} Y_{i g},
	\end{equation}
	as well as its conditional probability of expression given TF activity,
	\begin{equation}
		\hat{P}(g \mid t) = \frac{1}{|\mathcal{A}_t|} \sum_{i \in \mathcal{A}_t} Y_{i g}.
	\end{equation}
	Intuitively, $\hat{P}(g)$ measures how frequently a gene is expressed overall, whereas $\hat{P}(g \mid t)$ measures how frequently it is expressed specifically in TF-active cells.

	Gene--TF interactions were quantified using a pointwise mutual information (PMI)--like score,
	\begin{equation}
		\mathrm{PMI}(g,t) = \log \frac{\hat{P}(g \mid t)}{\hat{P}(g)}.
	\end{equation}
	Positive values indicate genes that are enriched in TF-active cells relative to their baseline expression frequency, while negative values indicate depletion. A small regularization constant was added to both probabilities to avoid numerical instabilities when probabilities approach zero.

	For each TF, PMI scores were standardized across genes by converting them to Z-scores,
	\begin{equation}
		Z_g = \frac{\mathrm{PMI}(g,t) - \mu_t}{\sigma_t},
	\end{equation}
	where $\mu_t$ and $\sigma_t$ denote the mean and standard deviation of PMI scores for TF $t$ across all analyzed genes. This normalization assumes that most genes are not specifically associated with the TF and therefore provides a natural background distribution.

	Two-sided $p$-values were computed assuming a standard normal distribution, and multiple testing correction was performed using the Benjamini--Hochberg procedure. Genes with false discovery rate--adjusted $q$-values below 0.05 and positive Z-scores were retained as significant TF-associated genes. For each TF, the activity quantile (between 0.5 and 0.95) and the minimum expression fraction (between 1\% and 20\%) were chosen on a $10\times10$ grid to maximize the number of genes with FDR-adjusted $q$-values below 0.05. Because these parameters are selected after testing, the nominal FDR should be regarded as approximate.

	To assess biological relevance, predicted TF--gene associations were compared against independently curated protein--protein interaction data from the STRING database \cite{szklarczyk2023string}. Interactions supported by STRING (version 12.5, combined score of at least 400) were considered validated, providing an external benchmark that is independent of the single-cell expression data. Results are reported for protein-coding genes. For comparison with a foundation model, we computed TF--gene PMI values with TF-Sapiens following \cite{pearce2025cross}: $P(g\,|\,a)$ was obtained by single-step decoding after the assay token $a$ (10x 3$'$ v3) and $P(g\,|\,\mathrm{TF},a)$ by two-step decoding with the TF as the first gene token, genes with $P(g\,|\,a)<10^{-8}$ were removed, and the same Z-scoring, Benjamini--Hochberg correction, and STRING validation were applied. For each method, genes were ranked by their Z-score, and the percentage of STRING-supported genes among the 25, 50, and 100 top-ranked genes was averaged over the five TFs.

	\subsection*{Isomap and Euclidean--Geodesic Distance Analysis}

To characterize the geometric structure of the transcriptional data manifold, we compared linear and non-linear notions of distance using principal component analysis (PCA) and Isomap. PCA provides a linear approximation that preserves directions of maximal variance, whereas Isomap estimates distances along the manifold itself by approximating geodesic distances between samples.

Let $\mathcal{S} \in \mathbb{R}^{N \times G}$ denote the set of single-cell gene expression profiles. For the spermatogenesis datasets (Fig~\ref{fig8}E,~F and \nameref{S1_Text}), $\mathcal{S}$ contains the raw counts of up to 500 randomly chosen cells per cell type. For Tabula Sapiens 2.0 (Fig~\ref{fig8}G), $\mathcal{S}$ contains the scTOP-preprocessed profiles of up to 1,200 cells per tissue (at most 100 per cell type). In both cases distances are computed in the full gene space and $k = 5, 10, 15, 30, 45$. Isomap begins by constructing a $k$-nearest-neighbor graph $G=(V,E)$ on these points using Euclidean distances. Two samples $S_i$ and $S_j$ are connected by an edge if $S_j$ is among the $k$ nearest neighbors of $S_i$, with edge weight
\[
w_{ij} = \|S_i - S_j\|_2 .
\]
This graph provides a local approximation to the underlying manifold.

Geodesic distances on the manifold are then approximated by shortest-path distances on the graph. Specifically, for any pair of samples $(i,j)$, the geodesic distance $d_G(i,j)$ is defined as the minimal sum of edge weights along any path connecting $i$ and $j$ in $G$. This yields a distance matrix that reflects intrinsic manifold geometry rather than direct Euclidean separation.

To obtain a low-dimensional representation, classical multidimensional scaling (MDS) is applied to the squared geodesic distance matrix $D_G^2$. MDS finds coordinates $\{y_i\}\subset\mathbb{R}^d$ that minimize the stress function
\[
\sum_{i,j} \left( d_G(i,j)^2 - \|y_i - y_j\|_2^2 \right)^2 ,
\]
yielding an embedding that best preserves geodesic distances in a least-squares sense.

To quantify the extent to which non-linear geometry alters global relationships between cells, we compared pairwise Euclidean distances
\[
d_E(i,j) = \|S_i - S_j\|_2
\]
with the corresponding geodesic distances $d_G(i,j)$. We computed the Pearson correlation coefficient
\[
\rho = \mathrm{corr}\big(d_E(i,j),\, d_G(i,j)\big)
\]
across all sample pairs. Values of $\rho$ close to one indicate that Euclidean distances already approximate manifold geodesics, implying weak global curvature. Lower values signal stronger non-linear distortion.

\section*{Results}
\subsection*{Knowledge transfer across species}\label{sec2}
    \begin{figure}[!htbp]
     \centering
     \includegraphics[width=0.9\linewidth,height=0.78\textheight,keepaspectratio]{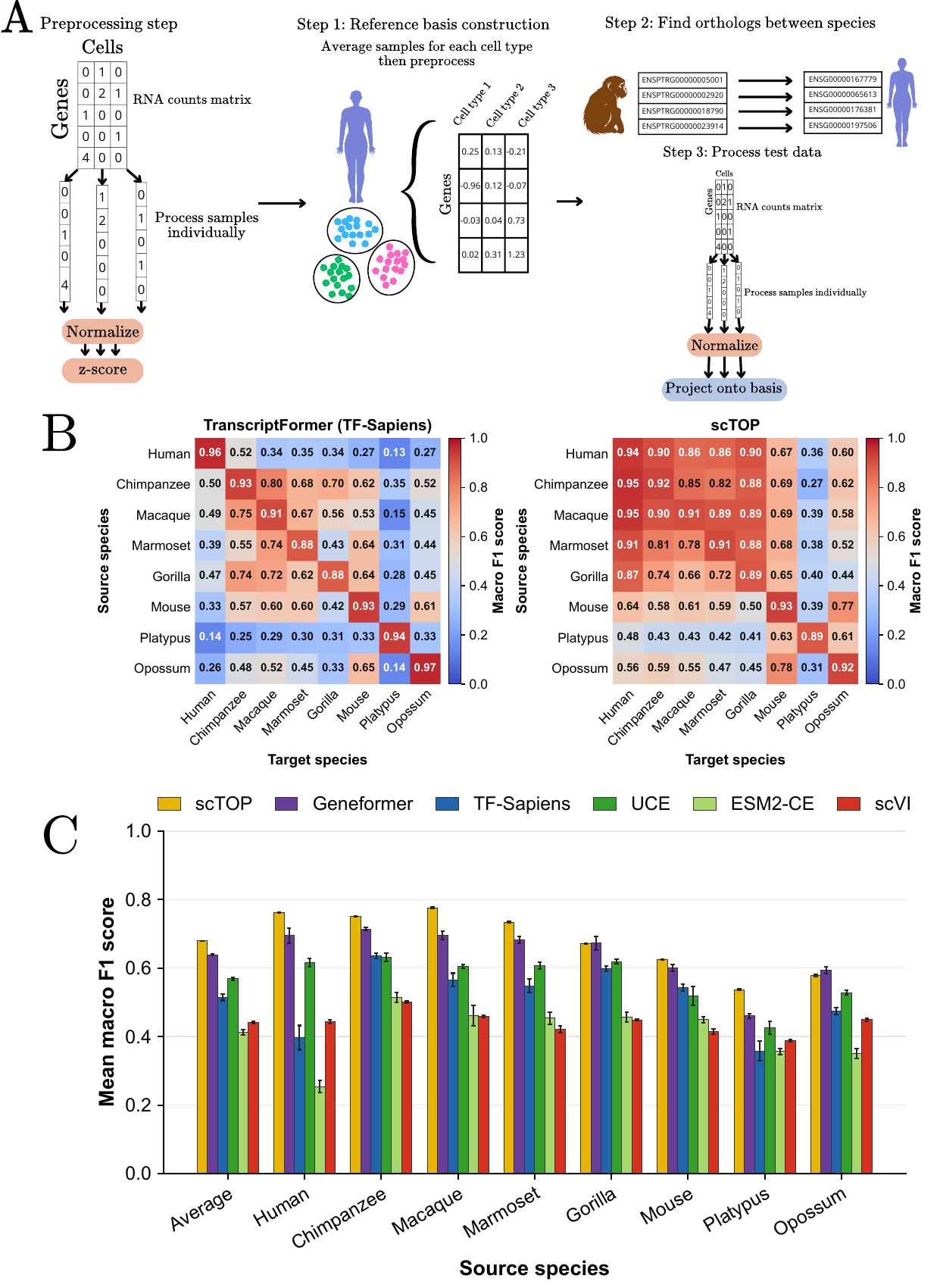}
     \caption{{\bf Cross-species transfer learning on novel organisms and cell types.} \textbf{A}. Pipeline used by \emph{parameter free}, linear algebra-based method scTOP  ~\cite{yampolskaya2023sctop} to perform cross-species annotation on the spermatogenesis dataset \cite{murat2023molecular,pearce2025cross}. \textbf{B}. Transfer
	matrices of macro F1 scores for testis cell type classification across mammals for the TranscriptFormer foundation model TF-Sapiens and scTOP performance. \textbf{C}. Comparison between scTOP and foundation models on the cross-species annotation task. Bars show the mean macro F1-score over the eight target species for each source species (and the overall average); error bars, standard deviation over five stratified 80-20 splits.}\label{fig3}
    \end{figure}

	Cross-species cell-type annotation provides a natural test of how biological identity is encoded in single-cell transcriptomic data. Because species differ in gene content, regulatory architecture, and evolutionary history, successful transfer requires isolating gene-expression programs that are conserved across evolution and robust to organism-specific variation.  For this reason, it is commonly assumed that successfully completing this task, especially on novel species and cell types absent from the training data, requires complex statistical models.

	For example, the authors of ~\cite{pearce2025cross} used  a `` transfer learning framework in which labels from one species (source species) [were] transferred and evaluated on embeddings from a different species (target species)'' to analyze a spermatogenesis dataset consisting of seven cell types (Sertoli cells, early spermatid, late spermatid, male germ line stem cell, somatic cell, spermatocyte, spermatogonium) and eight mammalian species (Human, Chimpanzee, Rhesus, Marmoset, Gorilla, Mouse, Platypus, Opossum). They interpreted the success of TranscriptFormer at this task as evidence that their embeddings successfully capture hard-to-learn evolutionary relationships between genes and cell types \cite{murat2023molecular}.

	We analyzed this task using TranscriptFormer-Sapiens and other foundation models. The results for TF-sapiens across species are shown in Fig~\ref{fig3}B, with larger scores (red) indicating better transfer learning. To better understand the difficulty of this task, we re-analyzed this dataset using scTOP \cite{yampolskaya2023sctop}. scTOP is a simple linear-algebra based method with {\it no free parameters} that constructs a reference basis for cell types of interests and classifies new cells by projecting their gene expression profiles onto this basis. The basic steps involved in scTOP are shown in Fig~\ref{fig3}A. A more detailed description can be found in the Methods, Ref.~\cite{yampolskaya2023sctop}, and the accompanying Python code at \href{https://github.com/Emergent-Behaviors-in-Biology/Linear-representations-for-scRNA-seq-data}{our Github repository}. Briefly,
	\begin{itemize}
		\item The gene expression of each cell is \emph{normalized to itself}, by converting mRNA counts to z-scores that reflect the rank ordering of genes within the cell (e.g. a gene whose expression percentile is 50th percentile is assigned a score $z=0$, a gene that is at the 84th percentile a $z=1$, etc). We have found that this normalizing procedure significantly eliminates batch effects because all cells are normalized independently  ~\cite{yampolskaya2023sctop}.
		\item A reference basis is created for cell types of interest (source). Due to the extremely noisy nature of scRNA seq data, this is done by creating normalized pseudo-bulk expression profiles by averaging the expression profiles of cells with the same source label.
		\item We assume that the number of genes is greater than the number of source cell types and hence the source cell types define a linear subspace of the full gene expression space.
		\item To classify a target cell, we calculate the (non-orthogonal) linear projection of the target cell on each of the source basis vectors. The target cell is labeled as the source cell type with the largest projection.
	\end{itemize}

	In order to compare cell types across species,  we restricted each dataset to orthologous genes and mapped all orthologs (Ensembl Compara, release 116) to their basis species counterparts for consistency (Fig~\ref{fig3}). This allows us to define a common coordinate system in gene space across species. This step is inherently lossy. While the human dataset retains 34,168 genes, other organisms are reduced to roughly 9,000 to 15,600 shared orthologs. We also restrict ourselves to eight mammalian species in the original spermatogenesis dataset  \cite{murat2023molecular} that could be directly accessed from the TranscriptFormer Github tutorial  using their built-in data processing function \cite{CZTutorial} (see Methods). For each species we used five stratified 80-20 train-test splits to construct the basis.

	For each foundation model, we extract the embeddings and use them to train a logistic regression with five different random seeds for a 80-20 train-test split. Following \cite{pearce2025cross}, each species is embedded with all of its genes (see Methods). Across all eight species, scTOP achieves a higher average macro F1-score than every foundation model (0.679, compared with 0.639 for Geneformer, the strongest foundation model, and 0.515 for TF-Sapiens), and it is the best representation for six of the eight source species (Fig~\ref{fig3}C). Importantly, this improvement persists even for evolutionarily distant species pairs, including transfers from platypus. These findings indicate that the conserved biological structure relevant for cross-species annotation is already strongly accessible through linear representations once appropriate normalization and orthology mappings are applied. These results suggest that the gene expression profile characterizing a cell type is extremely conserved. It further implies that the ``data manifold'' spanned by biologically-realized cell types in gene expression space can be well-approximated as a linear subspace.  A practical consequence of these observations is that one can reliably use a simple pipeline to annotate cells of one species using cells from another species.

	\subsection*{Identifying global  biological structure from data}\label{sec5}
	\begin{figure}[!htbp]
		\includegraphics[width=0.9\linewidth]{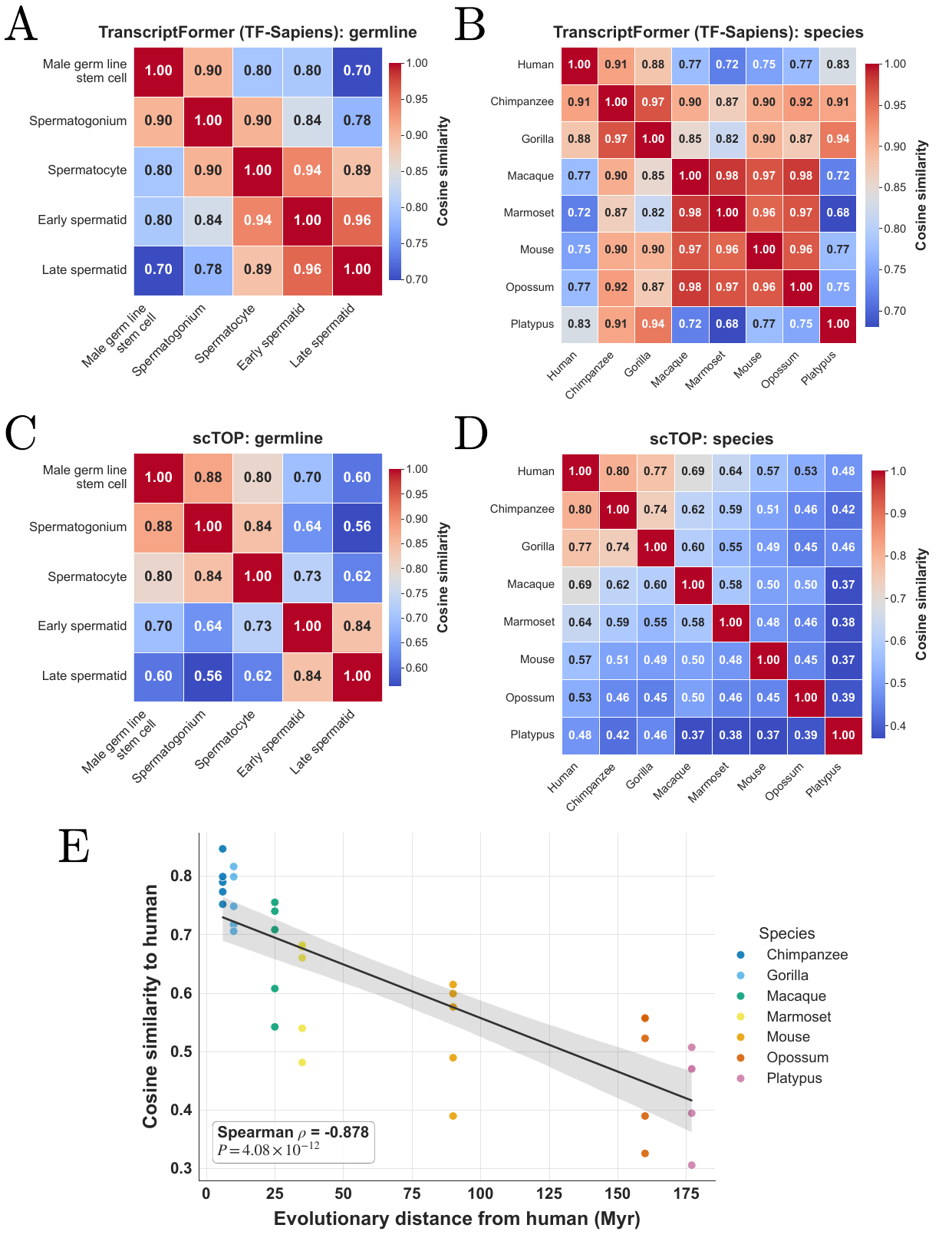}
		\caption{{\bf Biological context from data}. Cosine similarities for embeddings from TF-Sapiens and scTOP for different male germline developmental lineages and species. \textbf{A}. TranscriptFormer germline; \textbf{B}. TranscriptFormer species; \textbf{C}.  scTOP germline; and \textbf{D}. scTOP species. \textbf{E}. Cosine similarity from scTOP between humans and indicated species as a function of evolutionary distance.}\label{fig4}
	\end{figure}

	We also wanted to understand if simple vector representations of gene expression can capture global biological relationships between cell types and species. In \cite{pearce2025cross}, it was argued that one way of assessing this is by looking at the cosine similarity between embeddings for different cell types and species. In a good model,  the cosine similarity between cell types at similar developmental stages should be larger than the similarity between cell types at more distant stages. The similarity between species-level embeddings should also decay with evolutionary distance.

	Using the same dataset and evaluation protocol as \cite{pearce2025cross}, we analyzed representations generated by the scTOP preprocessing pipeline (Fig~\ref{fig3}A) and the TF-Sapiens embeddings. For each cell type, we constructed an average gene expression vector using pseudo-bulk gene counts as described in the last section and computed cosine similarities (Fig~\ref{fig4}A shows TF-Sapiens results and Fig~\ref{fig4}C shows scTOP results). As expected, within the male germline developmental lineage, we observe that cell types from successive developmental stages are more similar to each other than cell types from more distant stages. Comparing to TranscriptFormer embeddings, it is visually clear that early and late developmental stages are more distinct in linear representations than in the embeddings from foundation models.

	To compute the cosine similarity between different species, we averaged the gene expression profiles of cells from each species to compute a single representative species-level gene expression vector. In our pipeline, we restricted our analysis to gene orthologs and cell types shared by all species and for TF-Sapiens we used the embeddings from the model instead. As can be seen in Fig~\ref{fig4}D, our species-level gene expression vectors exhibit a much stronger evolutionary signal than embeddings from TranscriptFormer, with evolutionarily related species having much higher similarity than distant ones. The ability of appropriately normalized gene expression vectors to capture evolutionary relationships is also evident in  Fig~\ref{fig4}E, which shows the cosine similarity of different species as a function of evolutionary distance from humans. We find that cosine similarity and evolutionary distance are highly anti-correlated (Spearman's $\rho=-0.878$, two-sided $p=4.1\times10^{-12}$, $n=35$ species--cell-type pairs).

Collectively, our findings show that conserved developmental and evolutionary relationships between cell types and species are already quantitatively captured by properly normalized gene expression vectors and do not require the use of any statistical models or fitting.

	\subsection*{Tabula Sapiens cell type classification task}\label{sec3}
	\begin{figure}[!htbp]
		\centering
		\includegraphics[width=0.9\linewidth]{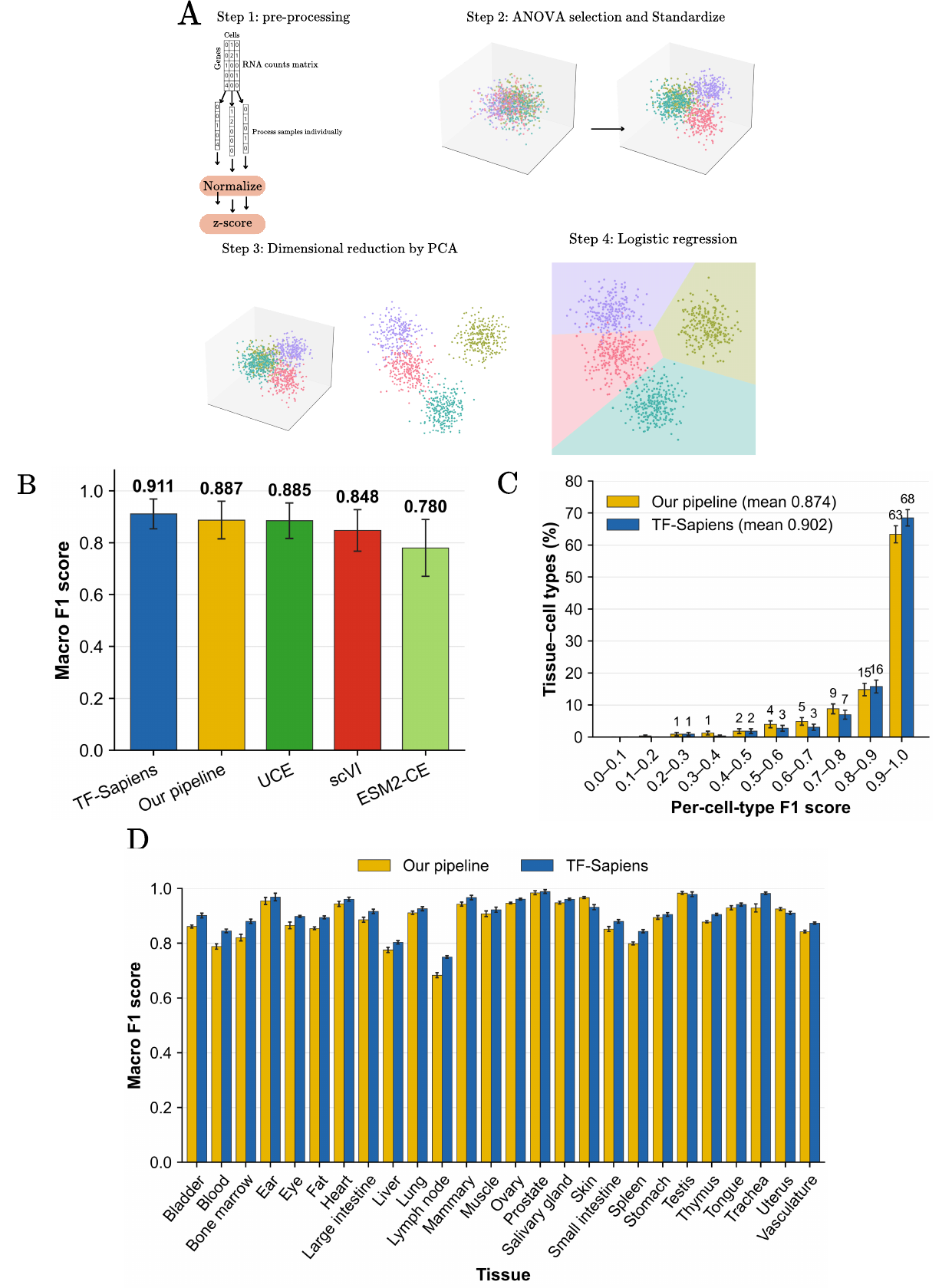}
		\caption{{\bf Tabula Sapiens 2.0 cell type classification task}. \textbf{A}: Overview of our pipeline. \textbf{B}: Average macro F1-scores across 26 tissues for our pipeline and foundation-model representations; error bars, standard deviation across tissues. \textbf{C}: Distribution of per-cell-type F1-scores for our pipeline and TF-Sapiens across all 330 tissue--cell-type combinations; whiskers, binomial standard error. \textbf{D}: Tissue by tissue comparison between TF-Sapiens and our pipeline; error bars, standard deviation across five stratified 80-20 splits.}\label{fig5}
	\end{figure}

	Human cell-type annotation provides a complementary setting to cross-species transfer for probing the nature of scRNA-seq data. Unlike evolutionary transfer, this task focuses on discriminating a large number of closely related cell types within the same organism, often in the presence of substantial technical noise and dropout. In ~\cite{pearce2025cross}, TranscriptFormer and other single-cell foundation models were benchmarked on this task using the Tabula Sapiens 2.0 dataset \cite{quake2025tabula}. This was done by training a tissue-level logistic regression classifier for cell identity on embeddings from each of the foundation models and then assessing the classifier on an independent test dataset of cells from the same tissue.

	We asked whether similar performance could be achieved with simpler pre-processing pipelines.  This prediction task involves distinguishing many more cell types than  the cross-species annotation task in the last section. For example, the immune tissue in Tabula Sapiens 2.0 dataset has more than 50 distinct cell types, many of which are closely related. As the number of cell types increases, the difficulty of the classification task increases dramatically, especially when the underlying data is noisy. This takes on added significance for the Tabula Sapiens 2.0 dataset, which in our experience, has much higher levels of technical noise and dropout than the spermatogenesis dataset analyzed above (see \nameref{S1_Text} for a discussion of batch effects in heart cell types as a typical example).

	In this noisier setting, raw linear projections alone are insufficient (see \nameref{S1_Text} for extended discussion). For this reason, we extended our pipeline by introducing two additional denoising steps, ANOVA-based gene selection and Principal Component Analysis (PCA), and replacing linear projections by a logistic regression-based classifier analogous to the one used to benchmark the single cell foundation models (see Methods for detailed explanation). These operations aim to isolate the most informative directions of variation while retaining simplicity and interpretability (see Fig~\ref{fig5}A).

	To perform our analysis, following ~\cite{pearce2025cross} we select all cell types from each tissue that have at least 100 cells and separate the data into training and test sets (80-20 ratio). We normalize the gene expressions of each cell independently as described above. We then perform an ANOVA on each tissue to select the 22000 genes that vary the most across cell types, followed by a standardization step in which each selected gene is scaled to zero mean and unit variance using the training cells. We then further denoise the data by performing a PCA and projecting the data onto the 220 most variable (whitened) PCA components (see methods section on hyperparameters search for a discussion on the choice of hyper-parameters). The ANOVA and PCA results from the training data sets are also applied to the test set to avoid information leakage. These de-noised cellular gene expression profiles play the same role in our pipelines as the cellular embeddings derived from the single-cell foundation models and serve as inputs to a logistic regression classifier (see Methods).

	Classification performance is evaluated using five different random seeds for the 80-20 train-test split. Fig~\ref{fig5}B shows the comparison between the average macro F1-scores across 26 tissues between our pipeline and foundation models. As done in \cite{pearce2025cross}, we evaluate 26 tissues: heart, ear, bladder, blood, bone marrow, eye, fat, large intestine, liver, lung, lymphnode, mammary, muscle, ovary, prostate, salivary gland, skin, small intestine, spleen, stomach, testis, thymus, tongue, trachea, uterus, and vasculature. Across 22 of 26 tissues, macro F1-scores are greater or equal to 0.8, indicating that a large fraction of the discriminative structure is captured by this representation (see Fig~\ref{fig5}D). Tissues with low scores, such as blood and lymph node, generally have many cell types that are closely related (e.g. different variant of T cells) and are also difficult to classify using single cell foundation models. For direct comparison with foundation models, we re-analyzed this dataset with the foundation models using the same splits and logistic regression on the embeddings. As shown in Fig~\ref{fig5}C and Fig~\ref{fig5}D, performance obtained with our pipeline closely matches (and in some cases exceeds) that for large foundation models, including TranscriptFormer. The mean macro F1-score across tissues is 0.887, compared to 0.911 for TranscriptFormer-Sapiens. Notably, the remaining performance gap is largely driven by a small subset of particularly difficult cell types with strong transcriptional similarity (see \nameref{S1_Text} for cells with lowest F1-scores and a more detailed cell type by cell type comparison).

	Fig~\ref{fig5}C shows the comparison between our pipeline and TF-sapiens distributions of macro F1-scores across individual cell types. More than half of all cell types achieve scores above 0.9, while the overall mean is dominated by a minority of hard-to-classify populations. This pattern suggests that, for most cell types, the relevant biological distinctions are already linearly accessible once noise is appropriately suppressed.

	Together, these results indicate that human cell-type annotation benchmarks primarily probe the effectiveness of denoising and feature selection, rather than the discovery of complex non-linear representations. Within the regime explored by current datasets, pipelines that isolate informative variance directions capture most of the structure required for accurate classification.

	\subsection*{Identifying cells infected by SARS-CoV-2}\label{sec4}
	\begin{figure}[!htbp]
		\includegraphics[width=\linewidth]{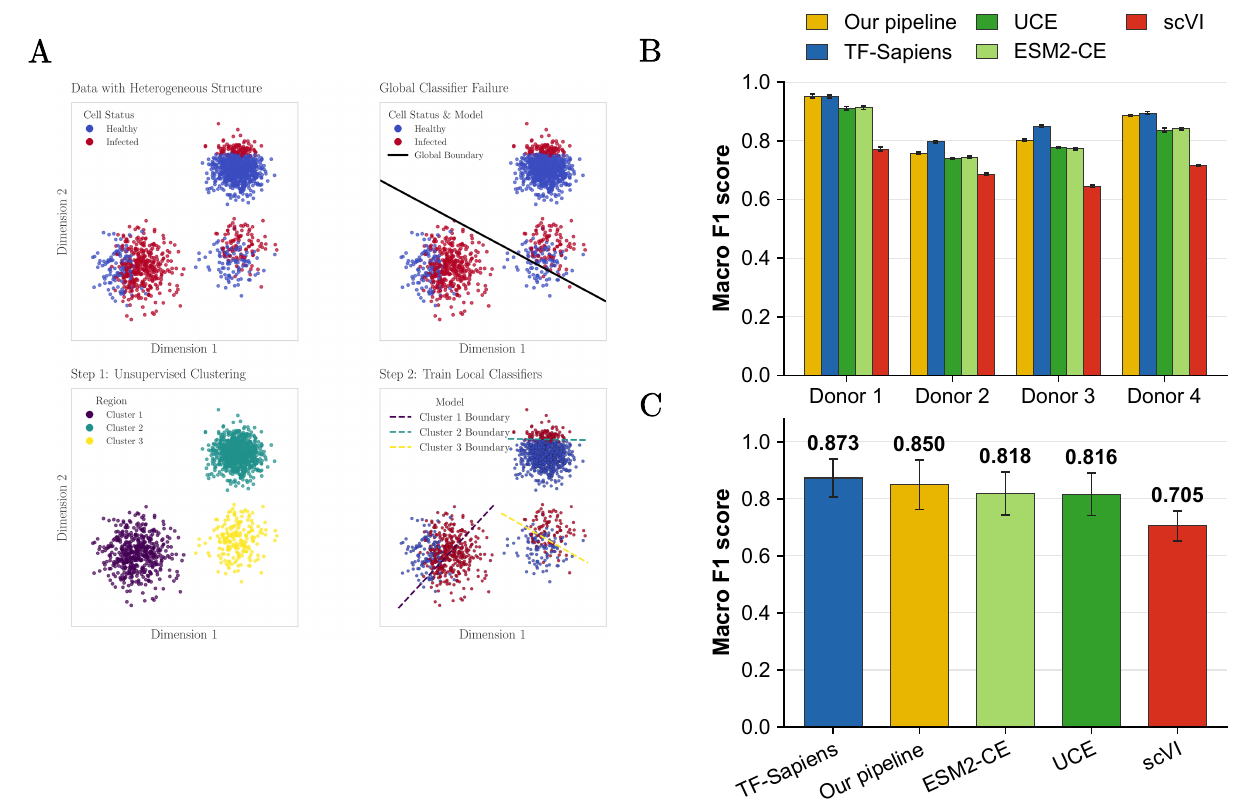}\\
		\caption{{\bf Identifying cells infected by SARS-CoV-2.} \textbf{A}. Schematic illustrating why local classifiers are necessary for this task. \textbf{B}. Comparison between our pipeline (yellow) and foundation models at classifying SARS-CoV-2 infected and uninfected cell from four distinct donors (data from \cite{wu2024interstitial}). \textbf{C}. Comparison of average disease state prediction F1 scores of uninfected and infected cells across all tissues and donors for our pipeline and foundation models.}\label{fig6}
	\end{figure}

	We next consider the task of classifying SARS-CoV-2–infected versus uninfected cells across multiple immune and lung cell types, using the dataset introduced in \cite{wu2024interstitial} and previously analyzed in \cite{pearce2025cross}. Distinguishing disease states within scRNA-seq data is a subtle task that presents new difficulties. Healthy and infected cells of the same cell type are highly correlated. There are also strong cell-type–specific class imbalances, with some populations containing nearly equal numbers of infected and uninfected cells and others being overwhelmingly composed of healthy cells.  These two challenges make it difficult to learn a single, global classifier that can distinguish infected and uninfected cells across all cell types (see Fig~\ref{fig6}a).

	To address these challenges, we augment the pipeline used in the last section to annotate human cell types (Fig~\ref{fig5}a) with an additional unsupervised clustering step (Fig~\ref{fig6}a). After normalizing the data, performing ANOVA gene selection and standardization, we used PCA (220 components) to perform dimensionality reduction and apply the Leiden clustering algorithm to partition the low-dimensional space into roughly 13 biologically coherent regions (clusters). Critically, this clustering operates without knowledge of health status and discovers natural subdivisions in the data based purely on transcriptional similarity.

	We then trained a separate logistic regression classifier within each cluster. This allows our pipeline to learn local classifiers that identify different context-dependent signatures of infection (see Fig~\ref{fig6}A). The key insight behind this strategy is that disease states are not globally separable, but locally distinguishable. Within a cluster of related cells, for example alveolar macrophages from similar tissue environments, the differences between healthy and infected states become pronounced. A classifier trained on this local structure can detect these differences without being confounded by the larger-scale variation across cell types and tissues.

	Our results demonstrate the power of this approach. For the SARS-CoV-2 dataset \cite{wu2024interstitial}, across all donors our method achieves a mean macro F1-score of 0.850,  exceeding the performance of most foundation models and achieving competitive performance to TF-sapiens (Fig~\ref{fig6}B,~C). These results further support a consistent interpretation across tasks: when biologically relevant signals are locally structured, extensions of linear pipelines that respect this locality are sufficient to recover disease-associated variation. Rather than requiring highly expressive global embeddings, accurate disease-state identification in this setting depends primarily on isolating the appropriate scale at which biological differences are expressed.

	\subsection*{Statistical structure of gene-gene and TF-Target relationships}
	\begin{figure}[!htbp]
		\centering
		\includegraphics[width=\linewidth,height=0.85\textheight,keepaspectratio]{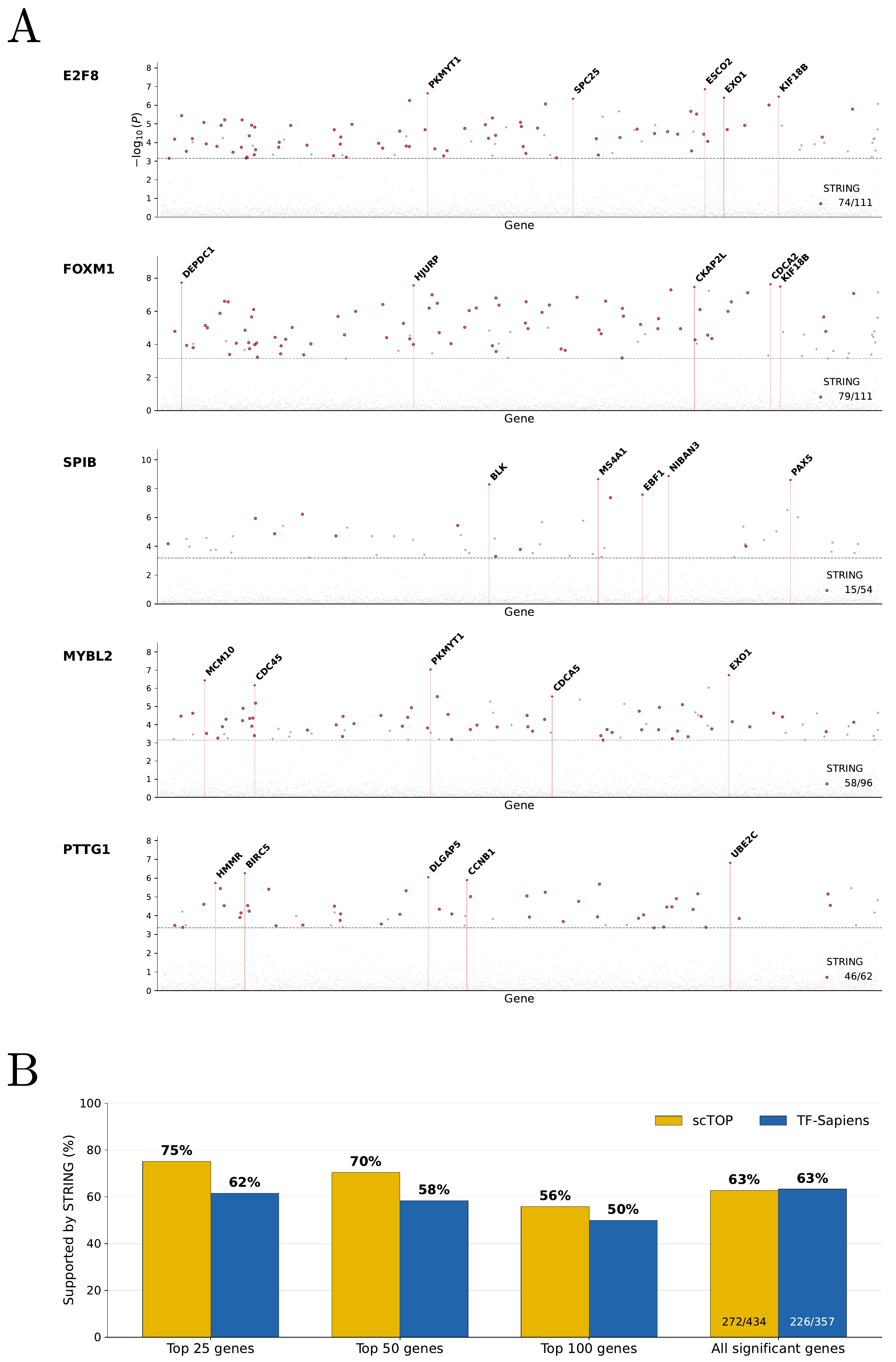}
		\caption{{\bf Validation of inferred TF-gene associations}. \textbf{A}. Each point represents   the Pointwise Mutual Information (PMI)-enrichment score between the indicated transcription factor (row name) and a protein coding gene. PMI scores are calculated using normalized scRNA-seq gene expression profiles from scTOP. Points lying above that dashed line indicate predicted interactions that are statistically significant after FDR correction, with points colored red indicating a protein-protein interactions that has been validated using the STRING database \cite{szklarczyk2023string}. \textbf{B}. Comparison with TranscriptFormer (TF-Sapiens): percentage of protein-coding genes supported by STRING among the 25, 50, and 100 top-ranked genes of each TF (mean over the five TFs) and among all significant associations (q $<$ 0.05).}\label{fig7}
	\end{figure}
	Single-cell foundation models such as TranscriptFormer have been used to identify interactions between transcription factors (TFs) and protein-coding genes. The approach works by using the model to generate synthetic gene expression profiles, then calculating how often a TF and target gene appear together in these generated profiles \cite{pearce2025cross}. This has been touted as a key advantage of foundation models: they can serve as a "virtual instrument" for probing gene regulatory relationships that would be difficult to study experimentally.

However, it remains unclear whether this capability is unique to foundation models or whether similar insights can be obtained directly from the data itself. To test this, we repeated the analysis from \cite{pearce2025cross}, but instead of using cellular embedding vectors from TranscriptFormer, we used the normalized gene expression profiles from scTOP as inputs. Briefly,  for each transcription factor, cells with high TF activity are identified, and the probability of observing each gene in these cells is compared to its marginal expression probability across the whole dataset. This defines a Pointwise Mutual Information (PMI)-inspired enrichment score that quantifies conditional dependence between TF activity and gene expression. Statistical significance is assessed using standard Z-scoring and false discovery rate correction (see Methods).

	We find that these empirically computed conditional statistics recover coherent and biologically meaningful TF–gene association patterns. When validated against independent protein–protein interaction evidence from the STRING database \cite{szklarczyk2023string}, 272 of the 434 significant associations (63\%) are supported by STRING, whereas each TF has at most 355 STRING partners among the roughly 20,000 protein-coding genes (Fig~\ref{fig7}A). Applying the same procedure to TranscriptFormer (TF-Sapiens) gives the same fraction (226 of 357, 63\%), but among the 25, 50, and 100 top-ranked genes for each TF, scTOP recovers on average 75\%, 70\%, and 56\% STRING-supported associations, compared with 62\%, 58\%, and 50\% for TranscriptFormer (Fig~\ref{fig7}B). This indicates that regulatory structure is already encoded in the observable conditional co-expression patterns of normalized scRNA-seq data, without requiring higher-order modeling or latent representations.

	\subsection*{Benchmark saturation and near-linear transcriptional geometry}

	\label{manifold}
	\begin{figure}[!htbp]
		\centering
		\includegraphics[width=0.9\linewidth,height=0.70\textheight,keepaspectratio]{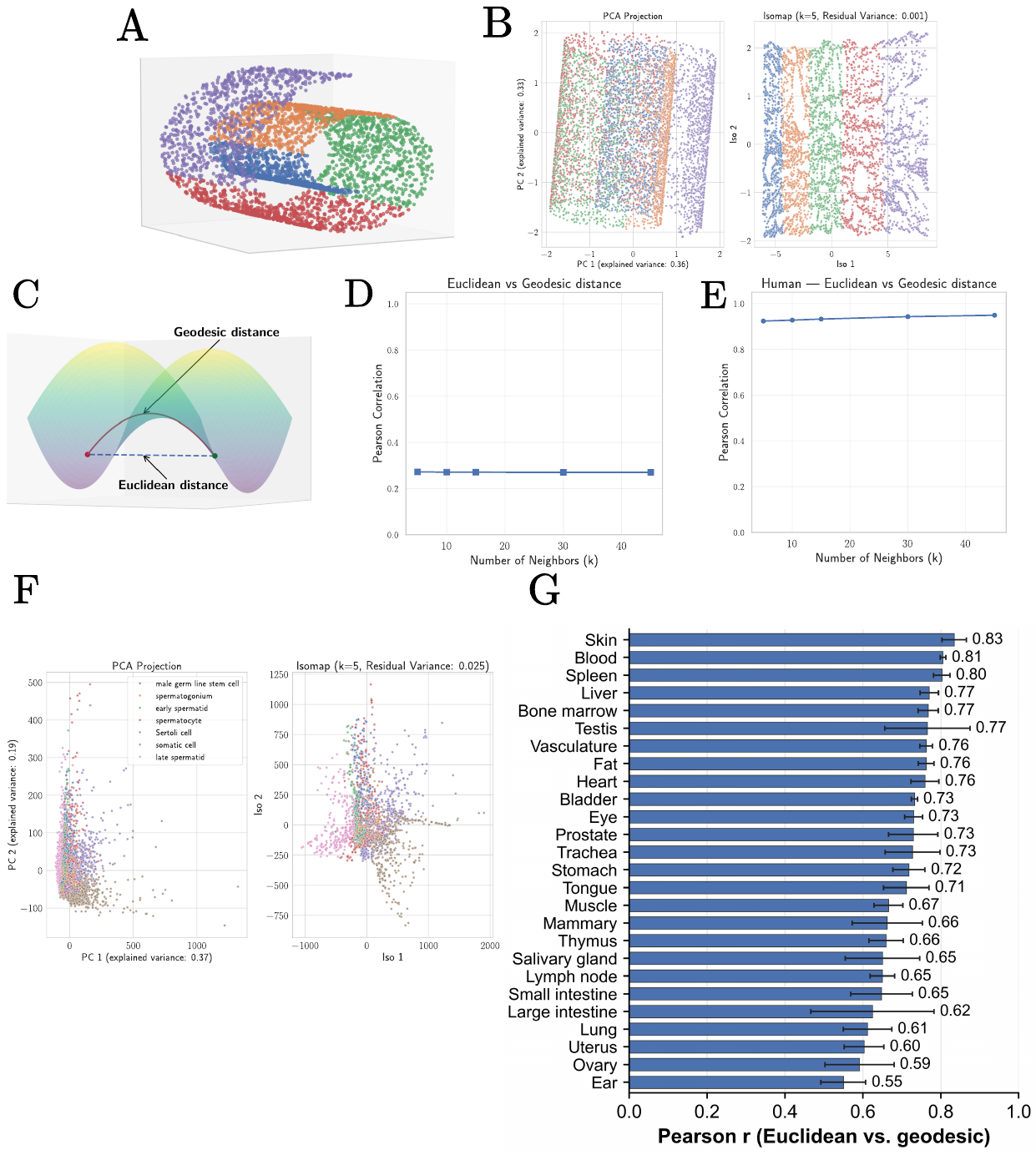}
		\caption{{\bf Detecting non-linear structure in data using geodesic distances from Isomap}. \textbf{A}. Swiss roll dataset in three dimensions. \textbf{B}. Two dimensional projections of the swiss roll dataset using Principal Component Analysis (PCA) and the non-linear manifold learning technique Isomap \cite{tenenbaum2000global} which uses the geodesic distance between points to create embeddings. \textbf{C}. Schematic illustrating difference between geodesic and Euclidean distance. \textbf{D}. Pearson correlation between Euclidean distances and geodesic distances between data points in the swiss roll dataset as a function of the Isomap hyper-parameter $k$ (number of nearest-neighbors). \textbf{E}. Correlation between geodesics and Euclidean distance for the human spermatogenesis dataset (raw counts) as a function of the Isomap hyper-parameter $k$ (number of nearest-neighbors). This correlation is higher than 0.9, providing strong evidence that the subspace spanned by the human spermatogenesis dataset is  approximately linear. \textbf{F}. Two dimensional embeddings of the human spermatogenesis dataset \cite{murat2023molecular,pearce2025cross} using Principal Component Analysis (PCA) and Isomap (raw counts).   \textbf{G}. Correlation between geodesics and Euclidean distance across all tissues studied in Tabula Sapiens 2.0, computed after scTOP preprocessing on up to 1,200 cells per tissue (at most 100 per cell type). Bars show the mean and error bars the standard deviation across $k = 5, 10, 15, 30, 45$.}
\label{fig8}
	\end{figure}

	 One of the surprising results of our analysis is the empirical observation that across diverse tasks -- cross-species transfer, within-species cell-type classification, disease-state prediction, and biological context extraction -- simple linear representations of cells in terms of normalized gene expression profiles can achieve SOTA or near SOTA performance comparable to extremely expressive single-cell foundation models. Here we provide a  potential explanation for this observation by thinking about gene expression datasets from a geometric viewpoint.

	  We view each cell as a point in gene space, with each gene is a different coordinate dimension (analogous to three-dimensional physical space). The position of a cell along each axis is just the normalized expression of that gene in the cell.  For scRNA-seq data where one simultaneously measures tens of thousands of genes, this space is extremely high-dimensional.  However, a key insight is that the effective space spanned by biologically realized cells is much lower-dimensional. The reason for this is that the set of biologically realized gene expression profiles across cells and species is a very small subset of all possible gene expression profiles.

	  In principle, this low-dimensional space could be highly structured. If this were the case, then representing cells would require extremely expressive models that can learn this complex structure. However, the empirical observation that  simple linear representations work well on downstream tasks led us to hypothesize that the submanifold corresponding to real cells is well-approximated by a \emph{linear subspace}.  To test this hypothesis, we created low-dimensional embeddings of our datasets using Isomap \cite{tenenbaum2000global}, an algorithm explicitly designed to model low-dimensional subspaces as curved manifolds, and compared them to low-dimensional embeddings obtained from  PCA, which explicitly assumes that data lives in a linear subspace.

	 For highly structured but low-dimensional data like the swiss roll dataset shown in Fig~\ref{fig8}A, these two algorithms produce embeddings with extremely different properties. In stark contrast to PCA which simply projects that data to a lower-dimensional representation, Isomap preserves geodesics distances, allowing it to identify the two-dimensional nature of the swiss roll dataset despite its high curvature Fig~\ref{fig8}B. This suggests that one reasonable proxy for assessing if a subspace is curved is to compare geodesic distances between data points obtained from Isomap embeddings with standard Euclidean distances in gene space (see Fig~\ref{fig8}C). As can be seen in Fig~\ref{fig8}D, for datasets like the swiss roll that occupy a highly curved submanifold, Euclidean and geodesic distances have a very low correlation. In contrast, for datasets that can be well approximated as a linear subspace this correlations should be close to one.

	To quantitatively assess whether biologically-realized cellular gene expression profiles are well approximated by a linear subspace, we used PCA and Isomap to create low-dimensional embeddings of the human spermatogenesis dataset analyzed above (using raw counts). This choice is inspired by our observation that this data is especially high quality, with relatively little technical noise (see  S1). As can be seen in Fig~\ref{fig8}F, the resulting low-dimensional embeddings are qualitatively similar. To more directly probe the geometry of this dataset, we compared the Euclidean distance between datapoints in the original gene space to the geodesic distance between points as calculated by Isomap. In contrast with the swiss roll dataset where these distances were very weakly correlated, for the spermatogenesis dataset this correlation is greater than $0.9$, providing strong evidence in favor of our hypothesis as shown in Fig~\ref{fig8}E (for an extended discussion on the other datasets of spermatogenesis see \nameref{S1_Text}). Moreover, we tested it on the scTOP-processed tissues of Tabula Sapiens 2.0 and show our results in Fig~\ref{fig8}G. We notice that the lowest value for correlation is for the ear tissue with mean of 0.55 across different values of $k$. Once again we see that the non-linearity across these datasets is modest.

	This geometric perspective helps explain why simple pipelines repeatedly match or exceed the performance of substantially more complex foundation models. When the space spanned by biological datasets is approximately linear, additional model expressivity is not translated into performance gains. It also underscores the need for better understanding the structure of scRNA-seq data to assess if complex foundation models are necessary given current datasets.

\section*{Discussion}
Over the last few years, vast amounts of resources have been dedicated to developing single cell foundation models using cell atlases. Advocates argue that these models have the potential to yield fundamentally new insights into gene regulation and cell identity \cite{bian2024scmulan, hao2024large, cui2024scgpt, yang2024genecompass, schaar2024nicheformer, kalfon2025scprint, rosen2023universal, pearce2025cross}. For example, the authors of \cite{pearce2025cross} claim that TranscriptFormer learns ``representations with emergent biological properties that surpass previous approaches'' and for this reason represents ``a paradigm shift in how we interact with cellular data by functioning as a virtual instrument for biological inquiry'' \cite{pearce2025cross}. Foundation models like TranscriptFormer are viewed by their proponents as starting points for the even more ambitious goal of creating AI-based virtual cell models that can serve ``as interactive knowledge bases capable of simulating complex cellular phenomena'' \cite{pearce2025cross}.

     Here, we present strong evidence suggesting that extremely simple methods with no or few parameters can achieve comparable and, in many cases, exceed the performance of single cell foundation models on downstream tasks such as cross-species annotation, human cell-type classification, disease-state identification, and biological context extraction.
 For example, we show that a pipeline combining gene orthology information with scTOP  ~\cite{yampolskaya2023sctop}, a simple \emph{parameter free} linear-algebra algorithm that represents cells by a single normalized gene expression vector, on average outperforms single cell foundation models on transfer learning cell types to novel species and cell fates (see Fig~\ref{fig3}). This is especially notable since it has been argued that the performance of single cell foundation models on out-of-distribution tasks ``provide[s] compelling evidence that broader evolutionary pretraining enhances biological generalization'' \cite{pearce2025cross}. The results presented make a compelling case that this is in fact not the case.

A common ingredient of these pipelines is careful, problem-specific preprocessing and model choice: each cell is normalized independently by the rank of its genes, informative genes are selected, and, for disease-state prediction, local classifiers are trained within transcriptionally defined clusters instead of a single global classifier. Because simple models are interpretable, such choices can be guided directly by the structure of the data. One key advantage of simple representations over foundation models is their high level of interpretability. For example, we have used scTOP  to great success to visualize developmental dynamics \cite{yampolskaya2023sctop} and  assess the fidelity of cells engineered using directed differentiation \cite{herriges2023durable, burgess2024generation, alber2026bidirectional}. These pipelines also naturally interface with promising theoretical ideas about cell fate grounded in dynamical systems and statistical physics \cite{rand2021geometry,saez2022statistically, jutras2020geometric, lang2014epigenetic,teschendorff2021statistical}. We have recently used scTOP to analyze cellular differentiation trajectories and identify signatures of different bifurcation classes directly from data \cite{yampolskaya2025finding}.

We emphasize that our present analysis is limited to current scRNA-seq atlases, datasets, and benchmarks. It is plausible that other, more complex, tasks may need the increased expressive power of foundation models, for example models that go beyond single cell states to describe how cells change along developmental time, when the data are rich enough to support such representations. More complicated representations may also be necessary for multimodal datasets that combine multiple types of biological measurements. However, due to the noisy nature of high-throughput single cell experiment, even multimodal data may be amenable to simple representations after appropriate preprocessing. This highlights the need for developing simple methods that can be used to benchmark future multimodal foundation models.

	To understand why linear representations perform so well on scRNA-seq data, we examined the organization of single-cell data from a geometric perspective. The results show that while scRNA-seq data do exhibit some degree of non-linearity, these effects are modest. As a result, performance saturates rapidly once noise is suppressed and the dominant sources of variation are isolated. In this setting, increasing model expressivity yields little additional benefit. Importantly, biological signals such as cell identity, disease signature, and tissue context—often thought to require complex representations—are already recoverable through simple linear transformations. This suggests that current foundation models primarily function as sophisticated denoising procedures rather than tools that uncover fundamentally new biological structure.

     This raises questions about why the performance of single cell foundation models fails to live up to the lofty expectations of their proponents, especially given the success of transformer-based protein language models at sequence-structure prediction. A key reason for this is that scRNA-seq data is fundamentally different from the DNA sequencing data used for protein modeling. In lieu of discrete, high quality protein sequences, single-cell transcriptomics produces sparse, noisy measurements of continuous variables rife with technical artifacts and biological variability. In this setting, the inductive biases and high expressivity of neural networks may actually work against them. The biology of cell identity is also significantly more complex than protein biophysics. Whereas it is well accepted that a protein's sequence contains the information needed to specify its structure (Anfinsen's principle), cell biology is highly contextual and emerges from the complex interplay between transcription, post-translational modification, signaling, and environmental cues.

     From this perspective, the field's rush toward ever-larger foundation models likely needs reassessment. Our results suggest that current scRNA-seq cell atlases simply do not contain the kind of deep structure that justifies the use of complex deep learning models. The potential cost of this misalignment is considerable because of the incredible computational resources needed to train foundation models. Training TranscriptFormer required processing 100+ million cells on 1000 H100 GPUs \cite{pearce2025cross}, an undertaking accessible only to well-funded institutions with a substantial GPU infrastructure. Even deploying trained models requires considerable computational resources and expertise well beyond that possessed by a typical biology lab. For example, extracting gene and cell embedding using TranscriptFormer requires an A100 GPU.

      In addition, the complexity of foundation models means that they are necessarily black boxes with limited interpretability. For this reason, methods like scTOP that employ simple representations of cellular gene expression profiles represent a promising alternative to complex, computationally-intensive foundation models.

\section*{Supporting information}

\paragraph*{S1 Text.}
\label{S1_Text}
{\bf Supplementary analyses.} Lowest-scoring cell types and per-cell-type comparison with TF-Sapiens; scTOP-only ablation on Tabula Sapiens 2.0; computational cost comparison; manifold analyses for additional datasets; analysis of technical noise in Tabula Sapiens 2.0; linear discriminant analysis.

\section*{Data availability}
	\textbf{Spermatogenesis datasets:} \textit{Homo sapiens} \href{https://www.ebi.ac.uk/biostudies/arrayexpress/studies/E-MTAB-11063}{E-MTAB-11063}, \textit{Pan troglodytes} \href{https://www.ebi.ac.uk/biostudies/arrayexpress/studies/E-MTAB-11064}{E-MTAB-11064}, \textit{Gorilla gorilla} \href{https://www.ebi.ac.uk/biostudies/arrayexpress/studies/E-MTAB-11065}{E-MTAB-11065}, \textit{Macaca mulatta} \href{https://www.ebi.ac.uk/biostudies/arrayexpress/studies/E-MTAB-11068}{E-MTAB-11068}, \textit{Callithrix jacchus} \href{https://www.ebi.ac.uk/biostudies/arrayexpress/studies/E-MTAB-11069}{E-MTAB-11069}, \textit{Mus musculus} \href{https://www.ebi.ac.uk/biostudies/arrayexpress/studies/E-MTAB-11071}{E-MTAB-11071}, \textit{Monodelphis domestica} \href{https://www.ebi.ac.uk/biostudies/arrayexpress/studies/E-MTAB-11072}{E-MTAB-11072}, \textit{Ornithorhynchus anatinus} \href{https://www.ebi.ac.uk/biostudies/arrayexpress/studies/E-MTAB-11070}{E-MTAB-11070}.\\
	\noindent\textbf{Tabula Sapiens 2.0:} Subset of \href{https://cellxgene.cziscience.com/collections/e5f58829-1a66-40b5-a624-9046778e74f5}{Tabula Sapiens}. Tabula Sapiens 2.0 comprises donors 17-31, and the tissues used were: heart, ear, bladder, blood, bone marrow, eye, fat, large intestine, liver, lung, lymphnode, mammary, muscle, ovary, prostate, salivary gland, skin, small intestine, spleen, stomach, testis, thymus, tongue, trachea, uterus, and vasculature.\\
	\noindent\textbf{COVID-19:} All COVID-19 datasets are available at \href{https://cellxgene.cziscience.com/collections/2a9a17c9-1f61-4877-b384-b8cd5ffa4085}{CELLXGENE}.

\section*{Code availability}
	The code used for each task is available in \href{https://github.com/Emergent-Behaviors-in-Biology/Linear-representations-for-scRNA-seq-data}{our GitHub repository}. For simple tasks, the code is made available as easy to run Jupyter notebooks. For longer tasks such as COVID-19 disease identification and Tabula Sapiens V2, we provide the appropriate Python .py files, together with the cluster scripts used to run all foundation-model baselines.

\section*{Acknowledgments}
We are grateful to Maria Yampolskaya for many useful conversations regarding many of the ideas presented here. We would also like to thank Darell Kotton, Laertis Ikonomou, Eitan Vilker, members of the Mehta group, and the CZI Theory group. We are also grateful to Yeongmin Kim for useful comments on the first version of the manuscript. Artificial intelligence (AI) tools were used to assist with code generation and minor editorial revisions, including spelling, grammar, and typographical corrections.

\section*{Authors contributions}
\textbf{Conceptualization:} Huan Souza, Pankaj Mehta\\
\noindent\textbf{Formal analysis:} Huan Souza\\
\noindent\textbf{Funding acquisition:} Pankaj Mehta\\
\noindent\textbf{Supervision:} Pankaj Mehta\\
\noindent\textbf{Writing - original draft:} Huan Souza, Pankaj Mehta\\
\noindent\textbf{Writing - review and editing:} Huan Souza, Pankaj Mehta\\
\nolinenumbers

%
%

\bibliography{citations}

{\Large
\textbf{S1 Text: Supplemental Information}
}

	\section{Worst cell types on our pipeline}\label{worst_cells_SI}
\begin{figure}[htbp]
	\centering
	 \includegraphics[width=\linewidth]{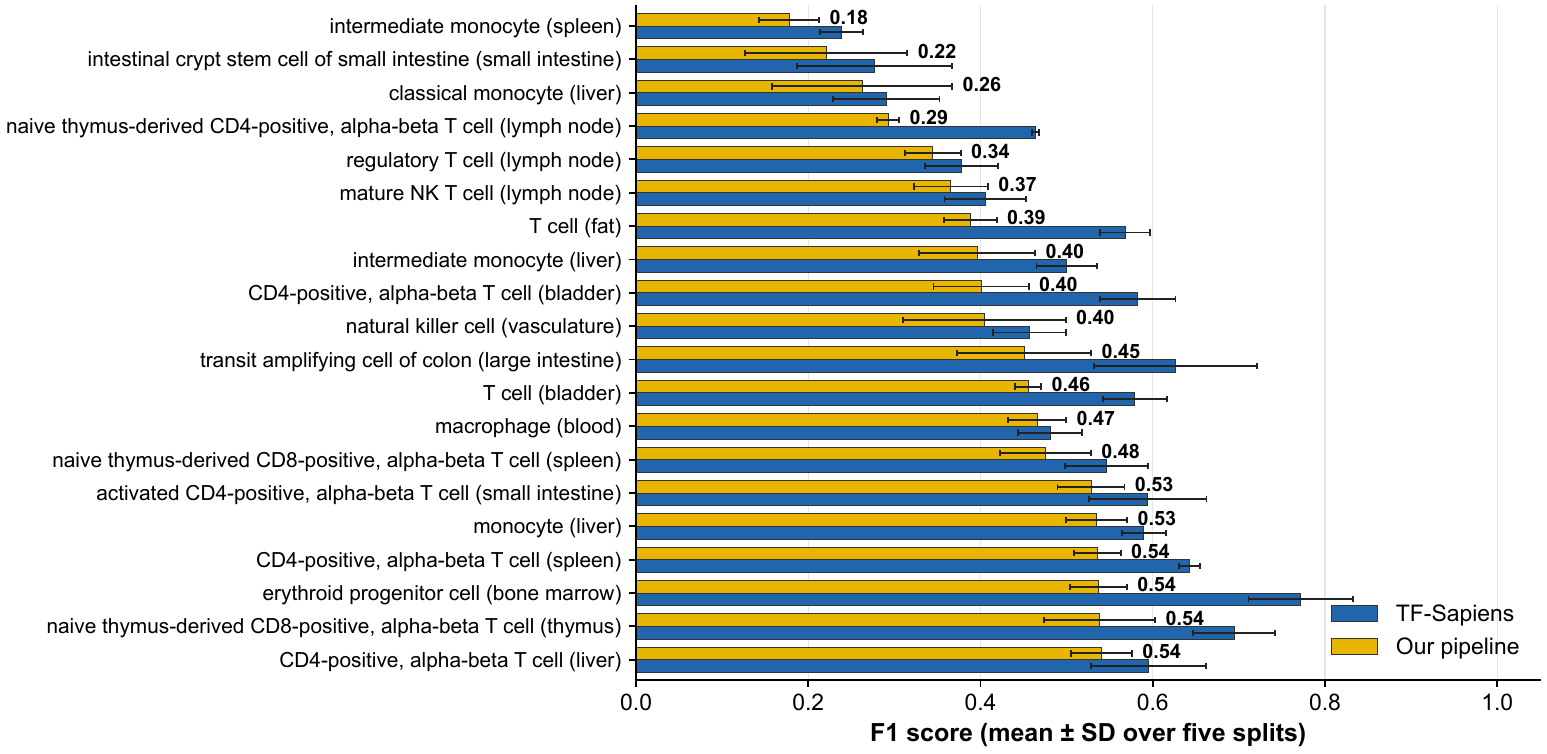}
	 \caption{{\bf Lowest-scoring cell types.} The 20 tissue--cell-type combinations with the lowest F1-score for our pipeline (mean $\pm$ standard deviation over five splits), with TF-Sapiens F1-scores on the same cells.}\label{worst}
	\end{figure}

\begin{figure}[htbp]
	\centering
	 \includegraphics[width=0.6\linewidth]{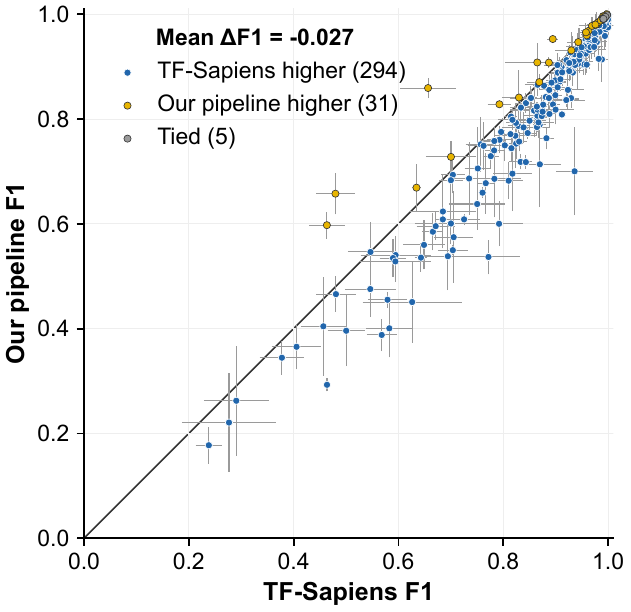}
	 \caption{{\bf Per-cell-type comparison with TF-Sapiens.} F1-score of our pipeline versus TF-Sapiens for each of the 330 tissue--cell-type combinations in Tabula Sapiens 2.0 (mean $\pm$ standard deviation over five splits). Our pipeline scores higher for 31 combinations, TF-Sapiens for 294, and 5 are tied; the mean difference is $-0.027$.}\label{scatter_SI}
	\end{figure}

	Similarly to what is reported in \cite{pearce2025cross}, we observe that the worst-performing cell types are in general T and NK cells. In Fig.~\ref{worst} we show the 20 tissue--cell-type combinations with the lowest F1-scores for our pipeline, together with the TF-Sapiens F1-scores on the same cells: 12 of the 20 are T/NK-lineage cells, and TF-Sapiens also scores low on most of them. Fig.~\ref{scatter_SI} compares the two pipelines for all 330 tissue--cell-type combinations.

	\section{Using scTOP only for Tabula Sapiens 2.0}\label{sctop_tabula}

	In the main text, we introduced a pipeline that doesn't use deep learning for annotating human cell
	types in Tabula Sapiens~2.0 that augments scTOP preprocessing with ANOVA-based gene selection, PCA,
	and linear classification. Here, we isolate the performance of scTOP alone in order to clarify
	both its strengths and its limitations on this task, and to explicitly motivate the denoising steps
	introduced in the main pipeline.

	scTOP represents each cell type by a single normalized pseudo-bulk expression vector and
	classifies individual cells by linear projection onto this reference basis. As shown throughout
	the main text, this representation is highly effective when the relevant biological structure is
	low-dimensional and robust, such as in cross-species annotation or developmental ordering.
	However, Tabula Sapiens~2.0 poses a fundamentally different challenge: each tissue contains
	a large number of closely related cell types, scRNA-seq measurements exhibit substantial
	technical noise and dropout, and within–cell-type transcriptional variability is often comparable
	to between–cell-type differences. In this regime, collapsing each cell type to a single mean
	vector discards discriminative information that is essential for fine-grained classification.

\begin{figure}[htbp]
		\centering
		\includegraphics[width=\linewidth]{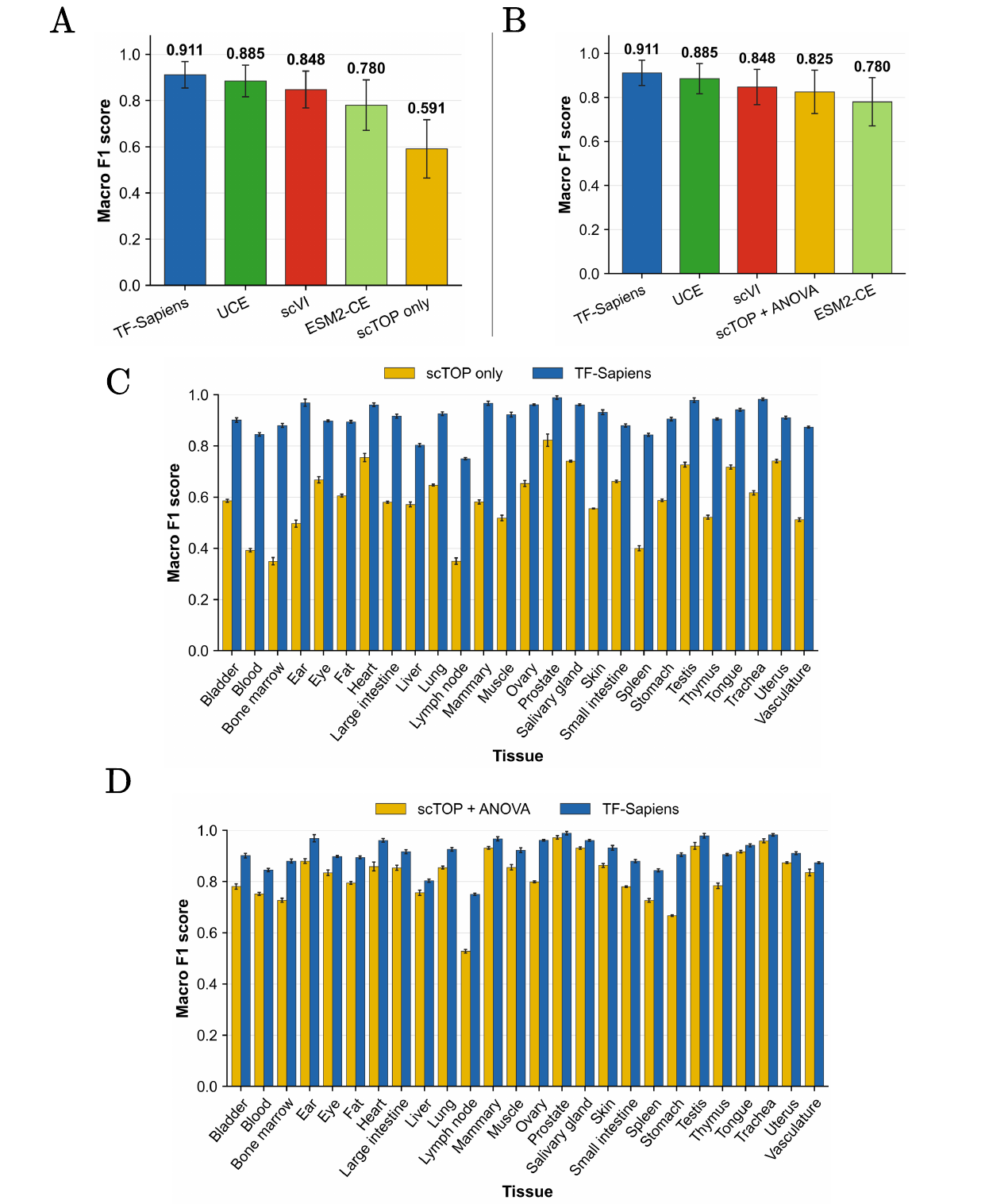}
		\caption{
			\textbf{scTOP performance on Tabula Sapiens~2.0.}
			\textbf{A}: Macro F1-score of scTOP alone compared to foundation-model baselines.
			\textbf{B}: Macro F1-score after adding ANOVA-based gene selection.
			\textbf{C}: Per-tissue macro F1-scores using scTOP alone.
			\textbf{D}: Per-tissue macro F1-scores for scTOP with ANOVA selection.
		}
		\label{sctop}
	\end{figure}

	Figure~\ref{sctop}A shows the macro F1-score obtained by scTOP alone across all tissues, evaluated
	using the same train–test splits and metrics as in the main text against the foundation models results. While scTOP performs well above chance and captures coarse cell-type structure,
	it consistently underperforms both foundation-model embeddings and the scTOP+ANOVA
	pipeline. This performance gap reflects a characteristic failure mode of centroid-based
	representations in noisy, high-dimensional settings: projections onto mean vectors are dominated
	by stochastic variability rather than cell-type–specific signal.

	The improvement obtained by introducing ANOVA-based gene selection (Fig.~\ref{sctop}B)
	follows directly from the conceptual framework developed in the main text. ANOVA explicitly
	filters out genes whose expression does not vary across cell types within a tissue, thereby
	removing directions in gene space dominated by technical noise and dropout. This operation
	does not introduce new information; instead, it reshapes the representation so that remaining
	dimensions preferentially encode inter–cell-type variability. As a result, even simple linear
	decision boundaries become effective once the signal-to-noise ratio is sufficiently enhanced.

	Figures~\ref{sctop}C and D show that this improvement is consistent across tissues. Importantly,
	the relative ordering of tissues by difficulty is roughly preserved, indicating that ANOVA selection
	acts as a uniform denoising step rather than a tissue-specific optimization.

\begin{figure}[htbp]
		\centering
		\includegraphics[width=\linewidth]{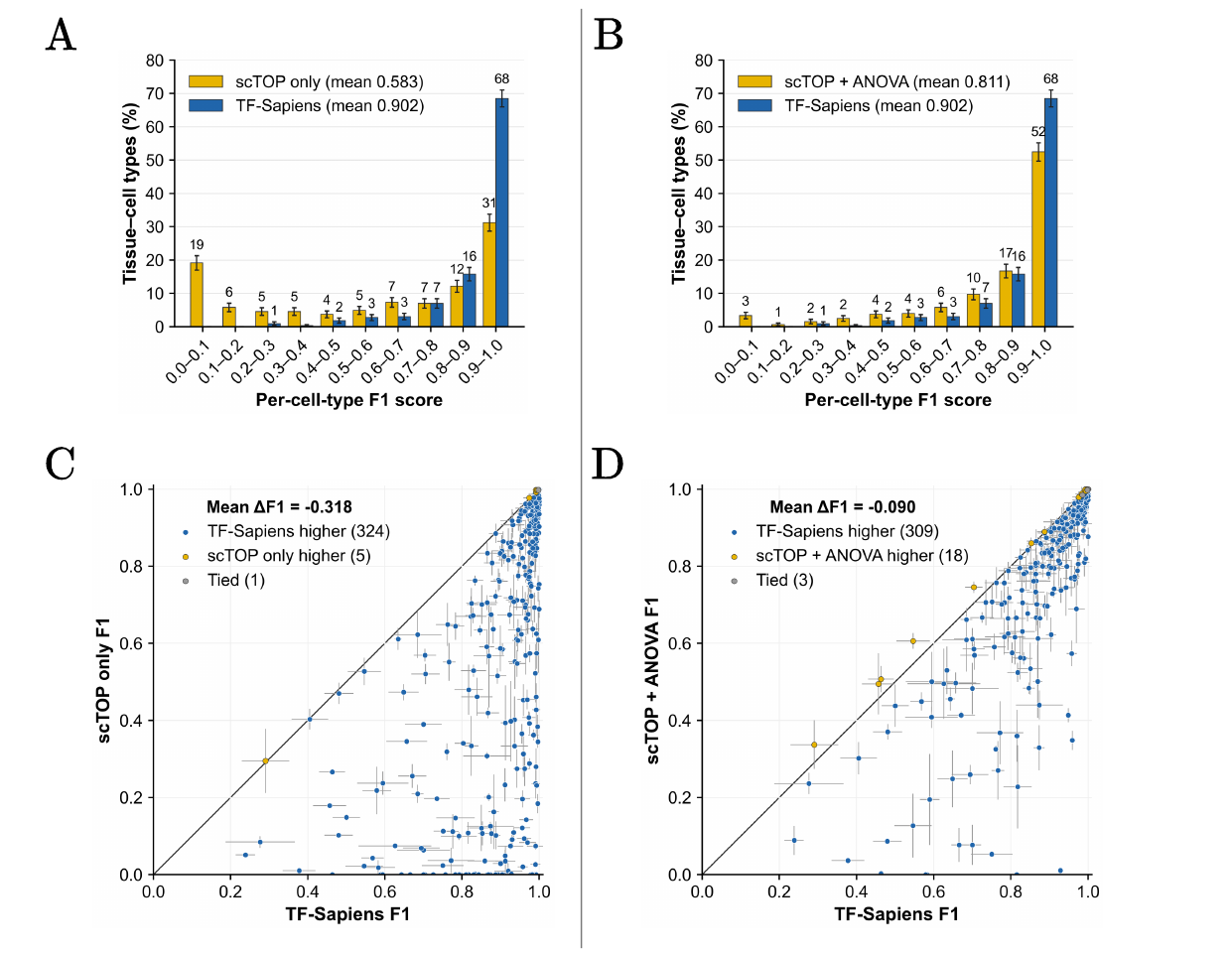}
		\caption{
			\textbf{Detailed cell-type-level performance.}
			\textbf{A}: scTOP alone histogram comparison of cell-type performance against TF-Sapiens.
			\textbf{B}: scTOP with ANOVA gene selection histogram comparison of cell-type performance against TF-Sapiens.
			\textbf{C}: Scatter plot of performance of scTOP alone pipeline compared with performance from TF-Sapiens
			\textbf{D}: Scatter plot of performance of scTOP with ANOVA gene selection pipeline compared with performance from TF-Sapiens
		}
		\label{sctop2}
	\end{figure}

	A more granular view of this effect is shown in Fig.~\ref{sctop2}, which shows per-cell-type performance. When using scTOP alone (Fig.~\ref{sctop2}A), confusion between
	transcriptionally similar cell types is pervasive, particularly for rare or weakly differentiated
	subpopulations. After ANOVA selection (Fig.~\ref{sctop2}B), these same cell-types exhibit
	sharply improved separation, despite the classifier and overall pipeline remaining entirely
	linear and having only a few hyperparameters.

	Taken together, these results demonstrate that scTOP’s limitations on Tabula Sapiens~2.0
	do not arise from the use of linear methods, but from representing highly noisy data using
	overly coarse summary statistics. The ANOVA selection step directly addresses this issue by
	aligning the representation with the structure of the classification task. In a subsequent SI
	section, we analyze the noise properties of Tabula Sapiens~2.0 in detail and show explicitly
	how dropout and within–cell-type variability degrade centroid-based representations.

	\section{Computational Cost, Throughput, and Scalability Analysis}\label{SI_comp}

	Here we provide a detailed and quantitative comparison of inference-time computational requirements for transformer-based cellular foundation models and the scTOP-based pipelines used throughout this work. We compare these approaches along multiple independent axes: floating-point operations (FLOPs), throughput (cells processed per second), scalability with dataset size, energy consumption, carbon footprint, and monetary cost.

	All comparisons are reported on a per-cell basis and are intentionally conservative, favoring foundation models whenever architectural details are not fully specified. Importantly, we focus exclusively on inference, which is the dominant operational regime for cell atlas annotation, cross-species transfer, and large-scale biological analysis.

	\subsubsection*{Algorithmic Decomposition of Inference Pipelines}

	\paragraph{scTOP.}
	Inference with scTOP consists of three strictly defined steps: (i) elementwise preprocessing of gene expression values, (ii) projection of the processed expression vector onto a precomputed reference basis of cell-type order parameters, and (iii) selection of the maximal projection score. The reference basis is constructed once and reused across all target cells; its cost is amortized and therefore excluded from per-cell inference cost, consistent with standard benchmarking practice.

	Let \(G\) denote the number of genes shared between reference and target datasets after ortholog mapping, and let \(K\) denote the number of reference cell types. The dominant operation is a dense matrix–vector multiplication of size \(G \times K\), yielding an exact inference-time complexity of
	\[
	\mathcal{O}(GK)
	\]
	floating-point operations (FLOPs) per cell.

	Across all experiments in this work, we observe \(G \in [9\times10^3, 6\times10^4]\) and \(K \in [3, 21]\), corresponding to
	\[
	5\times10^4 \;\text{to}\; 3\times10^6 \;\text{FLOPs per cell}.
	\]

	\paragraph{TranscriptFormer.}
	Inference with TranscriptFormer consists of a full forward pass through a large transformer-based encoder applied to each individual cell. As described in the original publication, TranscriptFormer models operate on gene tokens shared across species and compute contextualized cell embeddings via repeated blocks of multi-head self-attention and position-wise feedforward transformations. Inference therefore requires evaluating all transformer layers for every input cell, with no amortization across cells.

	The TranscriptFormer family includes three released models of increasing scale. The largest model (\textit{TF-Metazoa}) is trained on 112 million cells across twelve species and contains 444 million trainable parameters together with 633 million non-trainable parameters, for a total of over 1.07 billion parameters. Due to this scale, the processing of new cells for embedding generation effectively requires high-memory GPU hardware (e.g., A100-class accelerators).

	Let $n$ denote the number of gene tokens (up to a maximum sequence length of 2,047), $d = 2048$ the hidden dimension of the model, and $L = 12$ the number of transformer encoder layers. The dominant operations during inference arise from multi-head self-attention and dense linear projections. As for standard transformer architectures, the inference-time complexity per cell scales as
	\[
	\mathcal{O}\!\left(L(n^2 d + n d^2)\right),
	\]
	with the quadratic self-attention term $n^2 d$ dominating for biologically realistic values of $n$.

	Using the explicit model configuration reported in the paper ($L=12$, $d=2048$), inference with TranscriptFormer necessarily requires on the order of
	\[
	10^{11} \;\text{to}\; 10^{12} \;\text{floating-point operations per cell},
	\]
	depending on the specific model variant and the number of genes included. This estimate is a lower bound and excludes additional overhead from memory access, embedding aggregation, and downstream classification.

	As a consequence of this computational structure, TranscriptFormer inference throughput is limited to tens of cells per second even on modern GPUs. CPU-based inference, while theoretically possible, requires several seconds per cell due to the need to evaluate a billion-parameter transformer, resulting in throughput well below one cell per second and rendering CPU-based inference impractical for large-scale analyses.

	\begin{figure}[htbp]
		\centering
		\includegraphics[width=\linewidth]{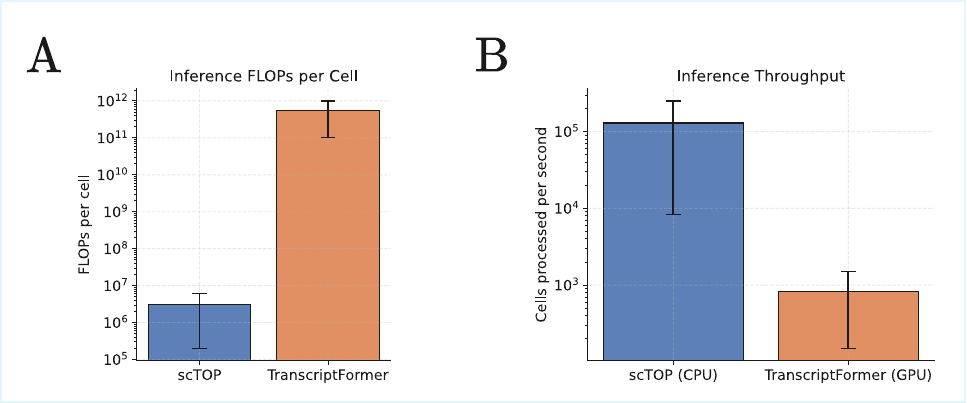}
		\caption{Comparison between the cost for extracting embeddings using TranscriptFormer models and pre-processing data using scTOP. The error bars indicate that the results are estimates and depends on the size of the model used (such as TF-exemplar, TF-sapiens, and TF-Metazoa). \textbf{A}: Floating point operations per-cell comparison. scTOP pre-processing is $10^5$ to $10^6$ times cheaper than extracting embeddings with TranscriptFormer. \textbf{B}: scTOP is $10$ to $10^3$ faster in processing cells than TranscriptFormer. Notice however that scTOP uses CPU while TranscriptFormer uses GPU, \textit{i.e.} is also much cheaper.}
		\label{comparison_fig}
	\end{figure}

	\subsubsection*{Inference Throughput: Cells Processed per Second}

	To translate FLOP counts into throughput, we estimate sustained effective compute rates under realistic conditions. We assume 50~GFLOPs/s for CPU execution and 3--30~TFLOPs/s for GPU execution, corresponding to typical effective performance rather than peak theoretical throughput.

	\paragraph{scTOP (CPU).}
	Under these assumptions, scTOP achieves inference throughput of approximately
	\[
	10^4\text{--}10^5 \;\text{cells per second}
	\]
	on commodity CPU hardware.

	\paragraph{TranscriptFormer (GPU).}
	On A100-class GPUs, TranscriptFormer achieves inference throughput on the order of
	\[
	10^1\text{--}10^2 \;\text{cells per second},
	\]
	depending on model variant and batch size. For reference, we measured 20--45 cells per second, including model loading, when embedding the spermatogenesis datasets with TF-Sapiens on a single GPU.

	\paragraph{TranscriptFormer (CPU).}
	Running TranscriptFormer inference on CPU is technically possible but computationally impractical. Using the same assumptions, CPU-based inference yields throughput of at most
	\[
	10^{-2}\text{--}10^{-1} \;\text{cells per second},
	\]
	corresponding to several seconds per cell. As a result, TranscriptFormer inference on CPU is slower than scTOP by approximately
	\[
	10^5\text{--}10^6 \;\text{fold},
	\]
	and is not viable for atlas-scale analyses.

	\subsubsection*{Scalability with Dataset Size}

	For a dataset containing \(N\) cells, total inference cost scales as
	\[
	\text{FLOPs}_{\text{scTOP}} \sim \mathcal{O}(NGK),
	\qquad
	\text{FLOPs}_{\text{TF}} \sim \mathcal{O}(NLn^2d).
	\]

	Because scTOP inference scales linearly with both dataset size and modest reference dimensionality, datasets containing millions of cells can be processed efficiently on commodity hardware. In contrast, TranscriptFormer inference scales superlinearly with gene count and requires specialized accelerators even for moderate dataset sizes.

	At the scale of modern cell atlases (\(N \sim 10^6\text{--}10^7\)), total inference FLOPs differ by a factor of approximately
	\[
	10^5\text{--}10^6.
	\]

	\subsubsection*{Energy Consumption and Carbon Footprint}

	To estimate energy usage, we adopt conservative efficiencies of 1.5~nJ/FLOP for CPU computation and 0.4~nJ/FLOP for GPU computation. Using a representative carbon intensity of 0.4~kg~CO\(_2\)/kWh, we estimate per-cell emissions.

	Under these assumptions, TranscriptFormer inference emits approximately
	\[
	10^5\text{--}10^6
	\]
	times more CO\(_2\) per processed cell than scTOP. This estimate excludes the substantially larger carbon footprint associated with training large foundation models, which would further widen this gap.

	\begin{figure}[htbp]
		\centering
		\includegraphics[width=0.5\linewidth]{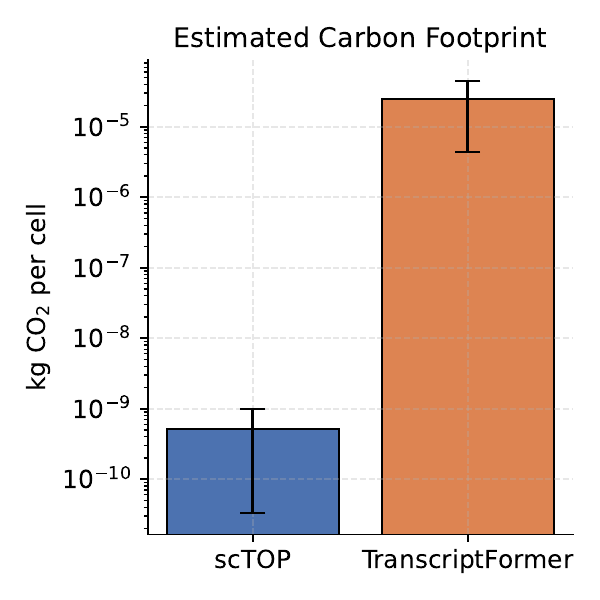}
		\caption{CO\(_2\) cost comparison between scTOP and TranscriptFormer during inference.}
		\label{co2_comparison}
	\end{figure}

	\subsubsection*{Monetary Cost of Inference}

	Assuming typical cloud pricing of \$3/hour for CPU instances and \$30/hour for A100-class GPUs, the per-cell monetary cost of TranscriptFormer inference exceeds that of scTOP by a factor of
	\[
	10^4\text{--}10^5,
	\]
	depending on batch size and deployment configuration. This cost differential scales linearly with dataset size and rapidly dominates total analysis cost for atlas-scale studies.

	\begin{figure}[htbp]
		\centering
		\includegraphics[width=0.5\linewidth]{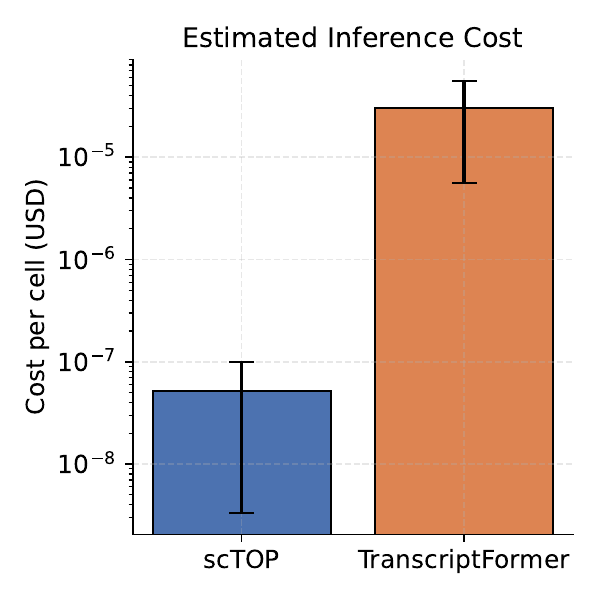}
		\caption{Monetary comparison between scTOP and TranscriptFormer during inference.}
		\label{money_comparison}
	\end{figure}

	\subsubsection*{Summary of Orders-of-Magnitude Differences}

	Across all metrics considered—inference-time FLOPs, throughput, scalability, energy consumption, carbon footprint, and monetary cost—scTOP outperforms TranscriptFormer by a factor of
	\[
	10^4 \;\text{to}\; 10^6
	\]
	at inference time while achieving comparable or superior predictive performance on the same benchmarks.

	These results demonstrate that the empirical success of large cellular foundation models cannot be interpreted independently of their extreme computational cost, and underscore the necessity of incorporating efficiency, scalability, and resource usage into benchmarking frameworks for biological representation learning.

	\section{Manifold analyses for more datasets}\label{manifold_SI}

	In this section we show the results of manifold analyses for more datasets beyond human spermatogenesis. We first show the results for other spermatogenesis datasets and then we show it for four Tabula Sapiens datasets chosen to represent different scores on our pipeline.
	\begin{figure}[htbp]
	 \includegraphics[width=\linewidth]{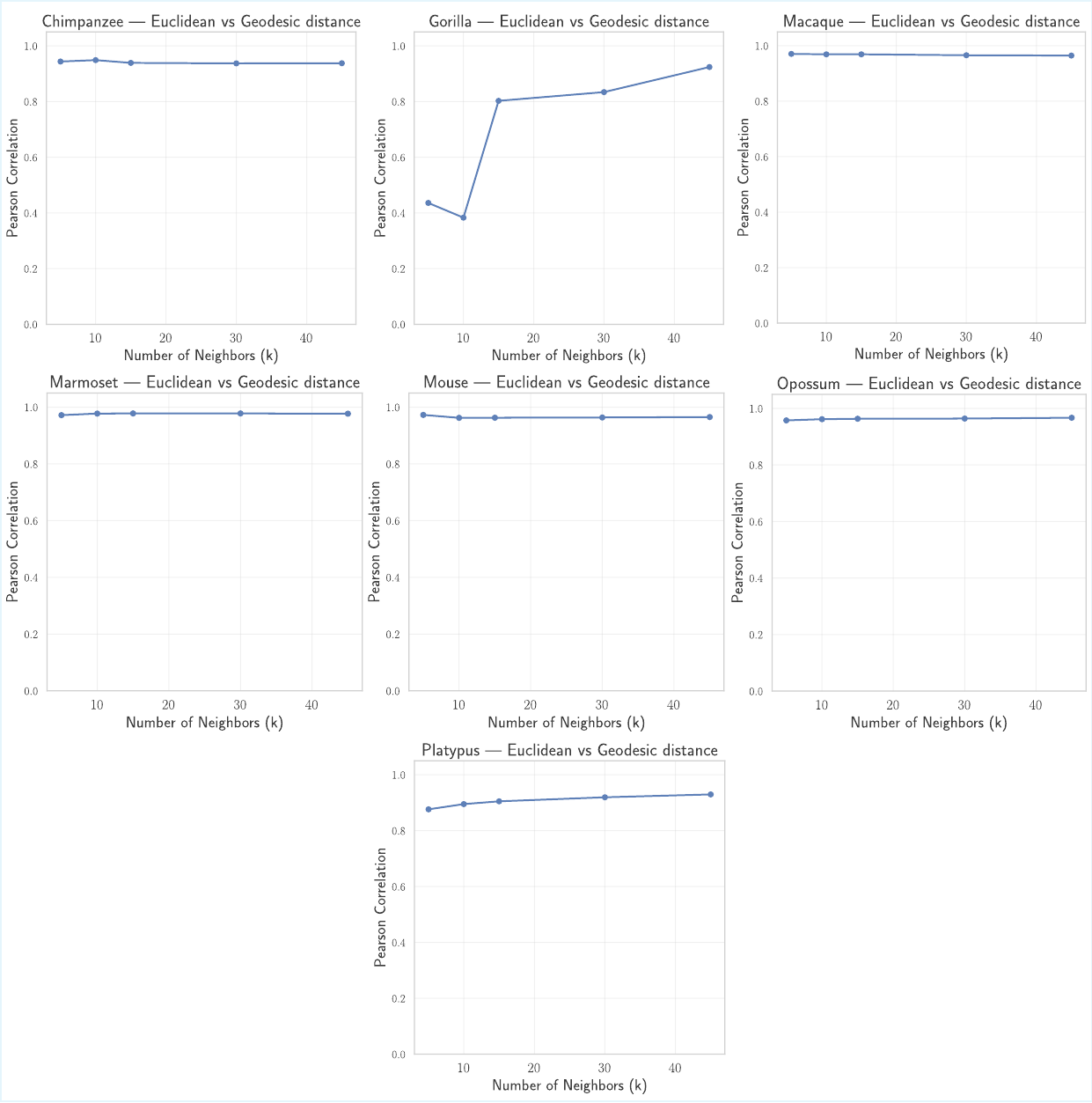}
	 \caption{Manifold analyses similar to the main text for the other seven spermatogenesis datasets (raw counts, up to 500 cells per cell type).}\label{species_manifold_SI}
	\end{figure}

	In Fig.~\ref{species_manifold_SI} we show the results of the same manifold analyses discussed in the main text but now for the other seven spermatogenesis dataset. We see that the results are similar to the human dataset, indicating that the near-linearity manifold feature remains true across datasets. The Gorilla dataset, on the other hand, yields unstable Isomap geodesics as a function of neighborhood size, suggesting that the kNN graph is not a robust manifold proxy for this dataset (e.g., disconnectedness at small k and/or short-circuit edges at larger k). Therefore the discrepancy reflects sensitivity of geodesic estimation to sampling/noise rather than a genuine change in tissue geometry.

	\begin{figure}[htbp]
	 \includegraphics[width=\linewidth]{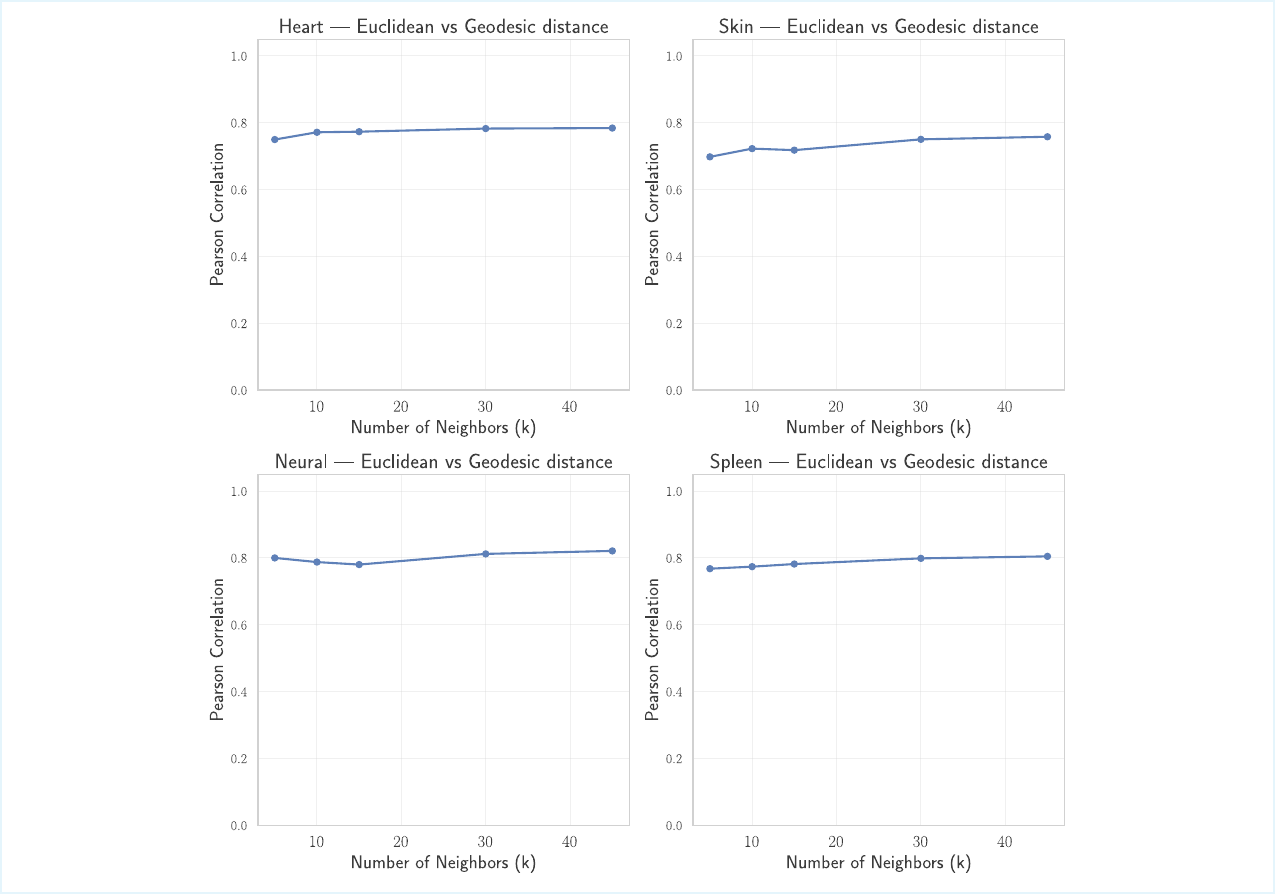}
	 \caption{Manifold analyses across four different tissues on Tabula Sapiens 2.0.}\label{tabula_manifold_SI}
	\end{figure}

	In Fig.~\ref{tabula_manifold_SI} we see that after using the process of scTOP the four tissues displayed have a correlation of around 0.8 between Euclidean distances and geodesics computed using Isomap. We observe that even for spleen tissue, that is among the lowest-scoring tissues (see Fig 5 in the main text) we obtain this high correlation. We interpret this as an evidence that the scRNA-seq data lives in a space that is not highly non-linear or has folding properties or holes. Moreover, most of the curvature and complications in this data arises from its technical noise and batch effects (see the next section for a discussion of technical noise in Tabula Sapiens 2.0).

	While Isomap provides a useful diagnostic for detecting non-linear structure by approximating geodesic distances on a neighborhood graph, its reliability depends critically on data quality and sampling density. In particular, Isomap assumes that local Euclidean neighborhoods faithfully approximate the underlying manifold; this assumption can break down in sparse, noisy, or heterogeneously sampled scRNA-seq datasets, where k-nearest-neighbor graphs may become disconnected or connected only through narrow bridges. In such regimes, estimated geodesic distances can become unstable with respect to the neighborhood size and may exaggerate apparent non-linearities that arise from technical noise, batch effects, or outliers rather than intrinsic biological structure. This sensitivity motivates using Isomap primarily as a qualitative probe of curvature, complemented by linear diagnostics and robustness checks as done throughout this work.

\section{Understanding levels of noise in Tabula Sapiens}\label{noise_SI}

\begin{figure}[htbp]
 \includegraphics[width=\linewidth]{
 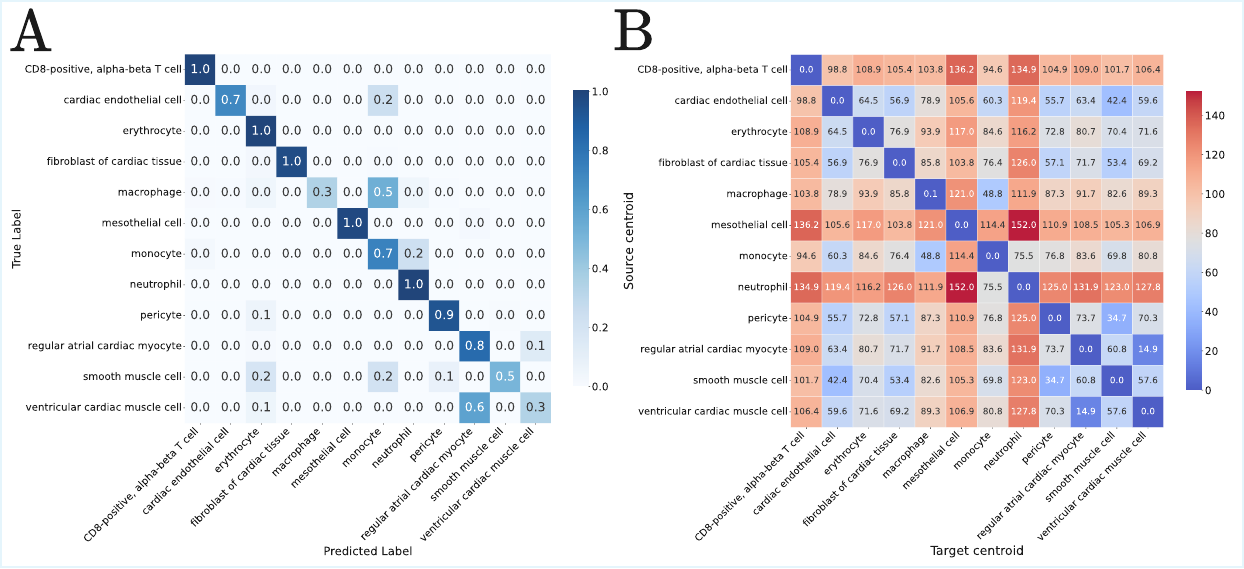}\caption{\textbf{A}: Confusion matrix for heart tissue. \textbf{B}: Distance between centroids across cell types.}\label{SI_noise_0}
\end{figure}

To further investigate the need of denoising in the Tabula Sapiens dataset, as an illustrative example we provide a more detailed analysis of technical noise in cells from the heart tissue. While using scTOP to classify cell types (see Fig.~\ref{SI_noise_0}A), we observed a surprising result: smooth muscle cells were mislabeled as distantly related cell types (erythrocyte, monocyte, and pericytes). While it is normal for cells that are biologically similar to get confused (e.g. atrial cardiac myocyte and ventricular cardiac muscle cell), misclassification involving cell types that are so developmentally dissimilar is rare. As can be seen in Fig.~\ref{SI_noise_0}B, the centroid of gene expression of smooth muscle cells in the training dataset is closer in gene space to, for example, cardiac endothelial cell and fibroblast of cardiac tissue than to erythrocyte and monocyte.

\begin{figure}[htbp]
 \includegraphics[width=\linewidth]{
 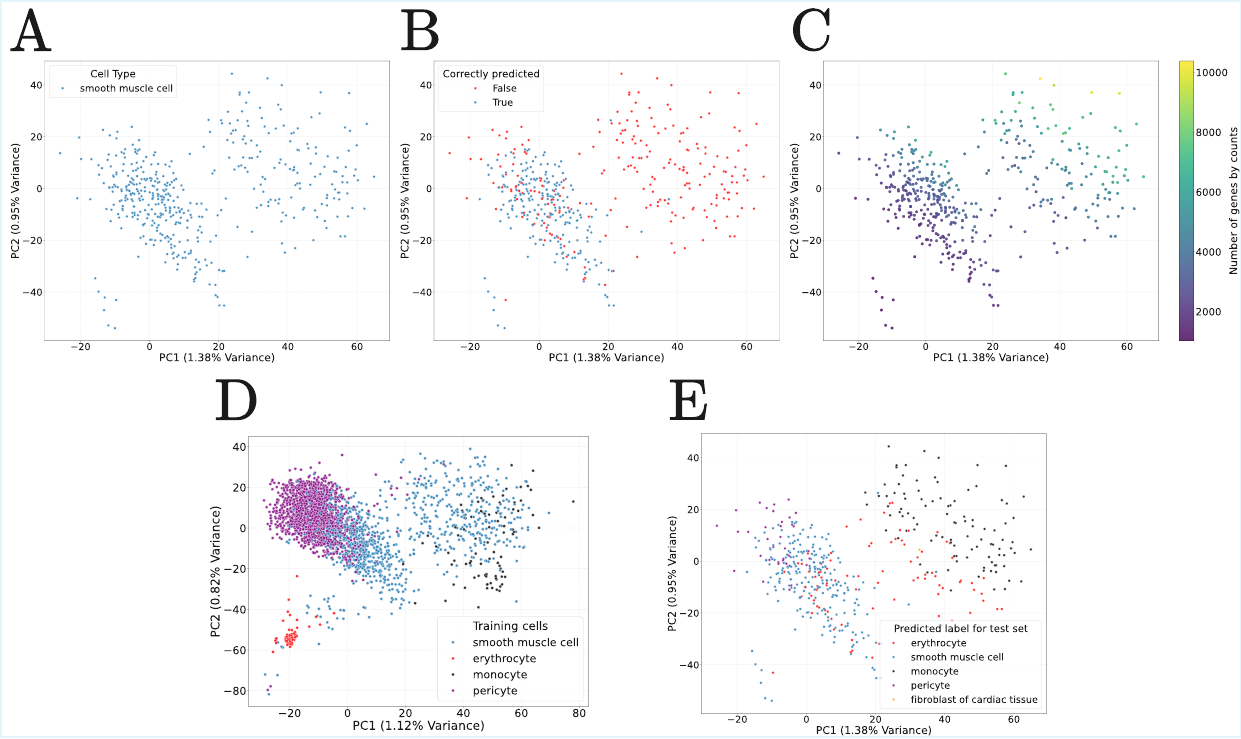}\caption{\textbf{PCA study of smooth muscle cells}. \textbf{A}: Smooth muscle cells PC1 and PC2. \textbf{B}: PCA for smooth muscle cells colored by correct and incorrect annotation. \textbf{C}: Smooth muscle cells colored by number of genes with at least one count. \textbf{D}: Training samples PC1 and PC2 for smooth muscle cells and cell types that it gets confused to. \textbf{E}: Smooth muscle cells PC1 and PC2 colored by predicted label for it.}\label{smooth_pca}
\end{figure}

To understand this behavior, we plotted the PC1 and PC2 for smooth cells in the test set (see Fig.~\ref{smooth_pca}A).  Even though these cells all share the same label, they form two clear and distinct clusters. Furthermore, one of these clusters is largely classified correctly by scTOP, while the other is assigned to the wrong cell type (Fig.~\ref{smooth_pca}B).  We noticed that these clusters differ in their technical noise. When we color the samples by the number of genes with at least one count in a cell, we see that the cluster on the left has a much higher dropout rate than the cluster on the right (see Fig.~\ref{smooth_pca}C). This pattern likely reflects technical artifacts arising from data collection and pre-processing rather than an intrinsic biology.

Fig.~\ref{smooth_pca}D shows a PCA of smooth muscle cells along with the cells for which they were confused (erythrocytes, monocytes, and pericytes). We notice that cells confused with pericyte indeed live closer to the pericyte cluster, cells confused with monocyte live closer to the monocyte cluster (see Fig.~\ref{smooth_pca}E). Meanwhile, cells labeled as erythrocyte are simply spread out over the whole PCA embeddings. Collectively, these observations suggest the source of scTOP's errors is likely technical noise.

These results also help explain  why ANOVA selection helps in annotation tasks. Since the differences in gene expression create two different clusters for a cell type, ANOVA acts as a denoiser of dimensions that create artificial differences between cells in the same cell type. Since it works by selecting genes that minimize variance within clusters and maximize variance between clusters, it effectively chooses genes that make each cell type cluster tighter while placing it as far as possible from the others.

\section{Linear Discriminant Analysis}

	In this section, we use Linear Discriminant Analysis (LDA) to further probe the linear separability of cell types. After using the full pipeline described in the Materials and methods section of the main text, we use LDA to understand if this is already enough to find meaningful directions that separate different classes.

	Unlike PCA, which is an unsupervised method that finds the axes of maximum total variance, LDA is a supervised method that finds the axes (linear discriminants) that best separate the known classes. It does this by maximizing the ratio of between-class variance to within-class variance. We apply LDA to the $d$-dimensional $\mathbf{z}_i$ vectors from the PCA step. For $K$ classes, we first compute the class means $\mathbf{m}_k$ and the overall mean $\mathbf{m}$:
	\[
	\mathbf{m}_k = \frac{1}{N_k} \sum_{i \in C_k} \mathbf{z}_i, \qquad \mathbf{m} = \frac{1}{N} \sum_{i=1}^{N} \mathbf{z}_i,
	\]
	where $C_k$ is the set of cells in class $k$ and $N_k = |C_k|$. The within-class scatter matrix $\mathbf{S}_W$ and between-class scatter matrix $\mathbf{S}_B$ are:
	\[
	\mathbf{S}_W = \sum_{k=1}^{K} \sum_{i \in C_k} (\mathbf{z}_i - \mathbf{m}_k)(\mathbf{z}_i - \mathbf{m}_k)^\top
	\]
	\[
	\mathbf{S}_B = \sum_{k=1}^{K} N_k (\mathbf{m}_k - \mathbf{m})(\mathbf{m}_k - \mathbf{m})^\top
	\]
	LDA finds the projection directions $\mathbf{w}_j$ by solving the generalized eigenvalue problem:
	\[
	\mathbf{S}_B \mathbf{w}_j = \lambda_j \mathbf{S}_W \mathbf{w}_j.
	\]
	The resulting projection vectors (eigenvectors) form a new basis $\mathbf{W}$ that projects the data into a new space of at most $K-1$ dimensions, where the classes are maximally separated.

\begin{figure}[htbp]
	 \includegraphics[width=\linewidth]{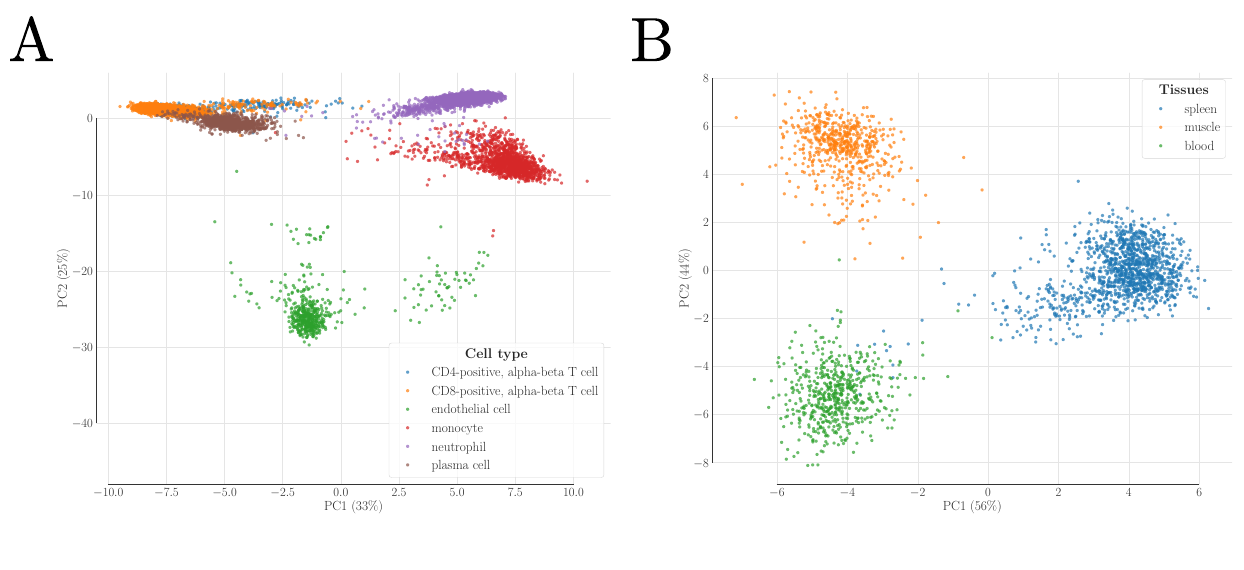}
	 \caption{\textbf{LDA analyses}. \textbf{A}: PC1 and PC2 projections of spleen tissue cells after using LDA show separability of different cell types after use of our pipeline. \textbf{B}: PC1 and PC2 of macrophages (spleen, muscle, and blood) show separation across tissues after the use of LDA.}\label{PCA_LDA}
	\end{figure}

	We perform the same analyses as described in \cite{pearce2025cross}. We train an LDA in donors from 1 to 16, then apply the trained model to donors 17-31. Importantly, this does not involve any deep learning. We first use this method to separate six cell types in spleen tissue (CD4+ alpha-beta T cell, CD8+ alpha-beta T cell, endothelial cell, monocyte, neutrophil, and plasma cell) as shown in Fig.~\ref{PCA_LDA}A. We see that we are able to get similar qualitative results as in \cite{pearce2025cross} using LDA. Moreover, we test if this pipeline can separate cells from a same cell type coming from different tissues. As shown in Fig.~\ref{PCA_LDA}B, we are able to distinguish different tissues for macrophages from spleen, muscle, and blood as done in \cite{pearce2025cross}.

\end{document}